\documentclass[]{aa}  

\usepackage{color} % Added to the A&A macro by AFL
\usepackage{graphicx}
\usepackage[varg]{txfonts}
\usepackage{lipsum}
\usepackage{amsmath}
\usepackage{natbib,twoopt}
\usepackage[breaklinks=true]{hyperref} %% to avoid \citeads line fills
\bibpunct{(}{)}{;}{a}{}{,} %% natbib format for A&A and ApJ
\makeatletter
\newcommandtwoopt{\citeads}[3][][]{\href{http://adsabs.harvard.edu/abs/#3}%
{\def\hyper@linkstart##1##2{}%
\let\hyper@linkend\@empty\citealp[#1][#2]{#3}}}
\newcommandtwoopt{\citepads}[3][][]{\href{http://adsabs.harvard.edu/abs/#3}%
{\def\hyper@linkstart##1##2{}%
\let\hyper@linkend\@empty\citep[#1][#2]{#3}}}
\newcommandtwoopt{\citetads}[3][][]{\href{http://adsabs.harvard.edu/abs/#3}%
{\def\hyper@linkstart##1##2{}%
\let\hyper@linkend\@empty\citet[#1][#2]{#3}}}
\newcommandtwoopt{\citeyearads}[3][][]%
{\href{http://adsabs.harvard.edu/abs/#3}
{\def\hyper@linkstart##1##2{}%
\let\hyper@linkend\@empty\citeyear[#1][#2]{#3}}}
\makeatother

\begin{document}

   \title{A model for spin-orbit commensurability and synchronous starspot activity in stars with close-by planets}

%   \subtitle{I. Overviewing the $\kappa$-mechanism}

   \author{A.~F.~Lanza
%          \inst{1}
          }
   \institute{INAF-Osservatorio Astrofisico di Catania, Via S.~Sofia, 78 - 95123 Catania, Italy    \\          
   \email{antonino.lanza@inaf.it}}
         
   \date{Received ... ; accepted ... }
\titlerunning{Spin-orbit commensurability in close-by planetary systems}
\authorrunning{A. F. Lanza}
% \abstract{}{}{}{}{} 
% 5 {} token are mandatory
 
  \abstract
   {%\LEt{ General notes: A.) You show a preference for US English grammar and spelling conventions, and I have edited accordingly throughout. B.) A\&A uses the past tense to describe specific methods used in a paper and the present tense to describe general methods as well as findings, including the findings of recent papers (within the past ten or so years). Kindly make any necessary changes (I have made some, but my edits are by no means exhaustive in this respect). See Sect. 6 of the language guide https://www.aanda.org/for-authors/language-editing/6-verb-tenses.}
   The rotation period of some planet-hosting stars appears to be in close commensurability with the orbital period of their close-by planets. In some cases, starspots rotating with a commensurable period have been detected, while the star displays latitudinal differential rotation.}
   {A model is proposed to interpret such a phenomenon based on the excitation of resonant oscillations in the interior magnetic field of the star by a component of the tidal potential with a very low frequency in the reference frame rotating with the star. }
   {A magnetic flux tube located in the overshoot layer of the star is assumed in order to study the excitation of the resonant oscillations in the magnetostrophic regime.  The model considers a planet on a circular oblique orbit, and the growth timescale of the oscillations is estimated. To keep the system in resonance with the exciting potential despite the variations in the magnetic field or tidal frequency, a self-regulating mechanism is proposed.}
   {The model is applied to ten systems and proves capable of accounting for the observed close commensurability in eight of them by assuming a magnetic field between $10^{2}$ and $10^{4}$~G. Systems with distant low-mass planets, such as AU~Mic and HAT-P-11, cannot be interpreted by the proposed model. }
   {Consequences for the spin-orbit evolution of the systems, including the dynamical tides and gyrochronology of planet-hosting stars, are discussed together with the effects on the chromospheric features produced by star-planet magnetic interactions.}


   \maketitle
%
%-------------------------------------------------------------------

\section{Introduction}

The majority of planets known to date have been discovered through the radial velocity or the transit methods, which favor the detection of planets close to their hosts \citep[e.g.,][]{Perryman18}. Therefore, tidal interactions between those planets and their host stars are remarkable and can affect both the orbital evolution of the planet and stellar rotation \citep[e.g.,][]{BolmontMathis16}. The overall distribution of the host rotation periods versus  the orbital periods of their innermost planets has been investigated  by \citet{McQuillanetal13} using Kepler space telescope data, but its interpretation is still a matter of debate \citep[e.g.,][]{TeitlerKoenigl14,LanzaShkolnik14,Ahuiretal21}. In addition to the difficulties in reproducing such a distribution, there are some intriguing cases where the orbital period of the planet, $P_{\rm orb}$, is in striking commensurability with the rotation period of its host star, $P_{\rm rot}$, as measured by means of the light modulation produced by its starspots or the rotation of the spots occulted during successive transits. 

One of the first such cases was that of Kepler-13A, which was reported by \citet{Szaboetal14} and shows $P_{\rm orb}/P_{\rm rot} = 5/3$, while the best studied case is probably that of Kepler-17, where $P_{\rm orb}/P_{\rm rot} = 1/8$ \citep{Desertetal11,BonomoLanza12,Bekyetal14,Lanzaetal19}. A list of systems that show this kind of commensurability is provided and discussed in Sect.~\ref{observations} and reveals that the integers appearing in the commensurability ratio can be remarkably different; this led such a phenomenon to be regarded as a bare coincidence in the vast ensemble of transiting planetary systems known to date. Yet, the closeness of the ratio to that of rather small integers $(\leq 8)$  suggests that the commensurability may be the result of an unknown physical process occurring in some systems (see, e.g., the discussion on Kepler-17 and HAT-P-11 by \citealt{Bekyetal14} or the case of AU~Mic recently investigated by \citealt{Szaboetal21}). 

In this work we propose a possible physical mechanism to account for the spin-orbit commensurability. It considers that the tidal potential produced by a close-by planet on an oblique or eccentric orbit has several components that oscillate with frequencies that are linear combinations of the stellar spin and orbital mean motion in a reference frame rotating with the star. When one such tidal  frequency comes close to the frequency of the magnetostrophic waves that can be excited in the magnetic fields in the stellar interior, a resonance can occur that can account for the observed phenomenology (Sect.~\ref{model}). The frequency of the magnetostrophic waves is much smaller than the spin frequency of the star, requiring a very small tidal frequency for their excitation in the reference frame rotating with the star. This implies that the commensurability between the stellar spin and the orbital motion must be tight in order to resonantly excite and maintain the magnetostrophic oscillations. On the other hand, once the oscillations are excited, they may be maintained by a self-regulating mechanism and may modify the spin-orbit evolution of the system by affecting the tidal torque and the stellar wind, which can contribute to maintaining the resonance for a rather long time interval. In Sect.~\ref{applications} we present specific models for the ten systems introduced in Sect.~\ref{observations} and discuss the advantages and drawbacks of the proposed model. In Sect.~\ref{discussion} we present our conclusions and mention some interesting applications of the presented model to other relevant problems in the field of star-planet interactions.

\section{Observations}
\label{observations}
\citet{Desertetal11} presented the intriguing case of Kepler-17, a system consisting of a solar-like star that shows spots that rotate with a period of 11.89 days, which is exactly equal to eight times that of the orbit of its hot Jupiter (see Table~\ref{table1}). Those spots were detected by the observations of their occultations by the planet during successive transits. The projected obliquity of Kepler-17 is smaller than $\pm 15^{\circ}$, while its orbit is circular \citep{Bonomoetal17}, thus simplifying the analysis of starspot occultations.  By co-adding the light curves of individual transits that occurred at epochs modulo eight planetary orbital periods, \citet{Desertetal11} found a remarkable repeatability of the pattern of occulted spots over a period of more than 250 days corresponding to 22 stellar rotations. The precision with which the commensurability is maintained can be conservatively estimated to be $0.75$\% from their Fig.~12 assuming that the recurrent spots do not deviate by more than $\pm \, 30^{\circ}$ over 22 rotations. The number of spots occulted during each transit ranges from two to three, suggesting that the number of starspots located along the occulted belt of the star is at least five or six. 

\citet{BonomoLanza12} performed a spot modeling of the out-of-transit light curve of Kepler-17 covering about 500 days and found that the large-scale spot pattern is organized into active longitudes with a mean rotation period of 12.01 days. This differs from the  rotation period of the occulted spots by $\sim 1-2$\% and suggests that only those occulted spots are in strict commensurability with the orbital motion of the planet. 

Using the times of light curve minima, \citet{Bekyetal14} reanalyzed the commensurability between the rotation of the occulted spots and the orbit of the planet and confirmed that the mean stellar rotation is different by about 1\% from  exact commensurability. A new analysis of the out-of-transit light curve of Kepler-17 covering about 1500 days was published by \citet{Lanzaetal19} who combined models of the out-of-transit light curve with those of the spots occulted during transits. They found that the star has an activity cycle of $400-600$ days characterized by an equatorward migration of the spots as in the Sun, while a shorter cycle of $\sim 48$~days, previously suggested by \citet{BonomoLanza12}, is intermittent and is not accompanied by any detectable migration of the spots in latitude.

Other systems showing a  commensurability between spot rotation and orbit of their planets are listed in Table~\ref{table1} where we report from the left to the right the name of the star, its mass $M$, radius $R$, effective temperature $T_{\rm eff}$, ratio of the orbit semimajor axis $a$ to the star radius, orbital eccentricity $e$, projected obliquity of the stellar spin axis $\lambda$, that is, the angle between the projections of the stellar spin and the orbital angular momentum on the plane of the sky,  mass of the planet $m_{\rm p}$, $P_{\rm rot}$, $P_{\rm orb}$, $P_{\rm rot}/P_{\rm orb}$, and relevant references. 

CoRoT-2 is particularly interesting because it shows a pattern with two main active longitudes rotating with a mean rotation period of 4.552 days that is in a 8:3 ratio with the orbital period of its hot Jupiter \citep{Lanzaetal09_C2}. In this star, the low-latitude spots occulted by the transiting hot Jupiter appear to rotate with a slightly shorter period of 4.48 days \citep{SilvaValioLanza11}. This difference with respect to Kepler-17 could be due to the faster rotation of CoRoT-2 that may produce the emergence of magnetic flux tubes coming from the base of the convection zone at higher latitudes outside the occulted belt owing to the stronger Coriolis force \citep[cf.][]{SchusslerSolanki92,Granzeretal00}. 
CoRoT-6 shows an orbital period shorter than the rotation period of the star and a starspot feature possibly synchronized with the orbital motion of its hot Jupiter \citep{Lanzaetal11}. CoRoT-4 is an F-type star showing starspots rotating synchronously with the orbit of its hot Jupiter that is not expected because of the weakness of the equilibrium tides given the relatively small mass of the planet, the relatively long orbital period of 9.2 days, and the thin convection zone of the host star \citep{Lanzaetal09}. 

The case of HAT-P-11 is remarkable because it shows a 6:1 commensurability within 0.034\% \citep{Bekyetal14}. The star has a high obliquity and the orbit is eccentric as in the case of Kepler-63 for which $P_{\rm rot}/P_{\rm orb} = (4.01 \pm 0.01)$:$7$ according to \citet{Sanchis-Ojedaetal13}. We also included:  WASP-107, which showed a ratio $P_{\rm rot}/P_{\rm orb} = (3.0 \pm 0.15)$ and has a high obliquity measured by means of spot occultations \citep{DaiWinn17}; AU~Mic, which has a high level of magnetic activity and a fast rotation corresponding to its very young age of $\sim 22$~Myr \citep{Szaboetal21}; and $\tau$~Boo,  where the mean stellar rotation is synchronized with the orbital motion of the planet.  

It is interesting to note that $\tau$~Boo has a differential rotation with an amplitude of $\sim 25$\% between the equator and the pole, but this does not prevent the possibility of finding synchronous activity in this system. More precisely,  \citet{Walkeretal08} claimed that $\tau$~Boo has a synchronized surface brightness inhomogeneity that maintained a fixed phase difference with the orbital motion of its hot Jupiter within $\sim 0.04$\% over a period of $\sim 400$~days. 

{In the present investigation, we do not consider spin-orbit commensurability in systems that consist of a star and a close-by brown dwarf, such as \object{CoRoT-33} \citep{Czimadiaetal15}, because we limit our application to companions in the planetary mass range. }

%%%%%%%%%%%%%%%%%%%%%%%%%%%%%%%%%%%%
\begin{table*}
\begin{small}
\caption{Systems showing evidence of spin-orbit commensurability.} % between the host star spin measured by means of starspots and the orbital mean motion. }
\begin{center}
\begin{tabular}{ccccccccccccc}
\hline
System & $M/M_{\odot}$ & $R/R_{\odot}$ &  $T_{\rm eff}$ & $a$  & $a/R$ & $e$ & $m_{\rm p}$ & $\lambda$ & $P_{\rm rot}$ & $P_{\rm orb}$ & $P_{\rm rot}$:$P_{\rm orb}$  & References \\
& ($M_{\odot}$) & ($R_{\odot}$) & (K) & (au) & & & ($M_{\rm J}$) & (deg) & (day) & (day) & ($m$:$n$) &   \\
\hline
\object{CoRoT-2} & 0.97 & 0.902 & 5625 & 0.0281 & 6.70 & 0.0 &  3.30 & 0.0 & 4.552 & 1.743 & 8:3 & 1, 2 \\
\object{CoRoT-4} & 1.16 & 1.17 & 6190 & 0.0902 & 17.36 & $<0.14$ & 0.703 & & 9.202 & 9.202 & 1:1 & 1, 3\\ 
\object{CoRoT-6} & 1.05 & 1.025 & 6090 & 0.0854 & 17.94 & $< 0.18$ & 2.95 & & 6.35 & 8.887 & 5:7 & 1, 4\\
\object{Kepler-13A} & 1.72 & 1.74 & 7650 & 0.0355 & 4.44  &  & $4.9-8.1$ & $23\pm 4$ & 1.046 & 1.764 & 3:5 & 5, 6, 7\\
\object{Kepler-17} & 1.16 & 1.05 & 5780 & 0.0268 & 5.48 & $ <0.02$ & 2.47 & $15\pm 15$ & 11.89 & 1.486 & 8:1 & 1, 8, 9\\
\object{Kepler-63} & 0.984 & 0.90 & 5575 & 0.080 & 19.1 & $<0.45$ & $ <0.38$ & $-110\pm 20$  & 5.401 & 9.424 & 3:5 & 10\\
\object{HAT-P-11} & 0.81 & 0.75 & 4780 & 0.053 & 15.6 & 0.218 & 0.0736 & $90\pm 30$ & 30.5 & 4.888 & 6:1 & 11, 12, 13, 14\\
\object{$\tau$ Boo} & 1.39 & 1.42 & 6400 & 0.049 & 7.40 & $< 0.011$ & 6.13 &  & 3.31 & 3.3124 & 1:1 & 15, 16 , 17\\
\object{WASP-107} & 0.69 & 0.66 & 4430 & 0.0558 & 18.16 &  0.0 & 0.12 & $40-140$ & 17.1 & 5.721 & 3:1 & 18 \\
\object{AU Mic} & 0.50 & 0.75 & 3700 & 0.0678 & 19.2 & 0.181 & 0.063 & $-3 \pm 10$  & 4.8367 & 8.463 & 4:7 & 19, 20 \\
\hline
\label{table1}
\end{tabular}
\end{center}
Note. References: 1: \citet{Bonomoetal17}; 2: \citet{Lanzaetal09_C2}; 3: \citet{Lanzaetal09}; 4: \citet{Lanzaetal11}; 5: \citet{Szaboetal14}; 6: \citet{Shporeretal14}; 7: \citet{Estevesetal15}; 8: \citet{Desertetal11}; 9: \citet{Lanzaetal19}; 10: \citet{Sanchis-Ojedaetal13}; 11: \citet{Bakosetal10}; 12: \citet{Sanchis-Ojedaetal11}; 13: \citet{Bekyetal14}; 14: \citet{Yeeetal18}; 15: \citet{Walkeretal08}; 16: \citet{Brogietal12}; 17: \citet{Borsaetal15}; 18: \citet{DaiWinn17}; 19: \citet{Szaboetal21}; 20: \citet{Caleetal21}. 
\end{small}
%\label{table1}
\end{table*}
%%%%%%%%%%%%%%%%%%%%%%%%%%%%%%%%%%

\section{Model}
\label{model}

\subsection{Tidal potential}
\label{tides}
In our model, the resonant oscillations of the interior stellar magnetic field are excited by the time varying tidal potential $\Psi$ due to a close-by planet of mass $m_{\rm p}$. Adopting a reference frame with the origin in the barycenter of the star, its polar axis along the spin axis of the star, and rotating with the stellar angular velocity $\Omega_{\rm s}=2\pi/P_{\rm rot}$, its expression is \citep[e.g.,][Sect.~2.1]{Ogilvie14}
\begin{equation}
\Psi  =  
\Re \left\{ \sum_{l=2}^{\infty} \sum_{m=-l}^{l} \sum_{n=-\infty}^{\infty} \Psi_{lmn}  \right\}, 
\label{tidal_potential}
\end{equation}
where $\Re \{z \}$ indicates the real part of a complex quantity $z$ and the Fourier components of the series development of the potential are given by
\begin{equation}
\Psi_{lmn} = \frac{Gm_{\rm p}}{a} A_{lmn} (e,i) \left( \frac{r}{a} \right)^{l} Y_{l}^{m} (\theta, \phi)  \exp[j(m\Omega_{\rm s}-n\Omega_{0}) t], 
\label{tidal_potential_comp}
\end{equation}
where $j = \sqrt{-1}$, $G$ is the gravitation constant, $a$ the semimajor axis of the relative orbit, $A_{lmn}(e,i)$ a coefficient depending on the eccentricity of the orbit $e$ and its obliquity $i$ \citep[that is, the angle between the stellar spin and the orbital angular momentum; see][]{Kaula61,MathisLePoncin09}, $r$ the distance from the center of the star of radius $R$, $Y_{l}^{m}(\theta, \phi)$ the spherical harmonic of degree $l$ and azimuthal order $m$ (with $|m| \leq l$) that is a function of the colatitude $\theta$ measured from the polar (spin) axis and the azimuthal coordinate $\phi$, $n$ is an integer specifying the temporal harmonic of the potential oscillations, and $\Omega_{0} = 2\pi/P_{\rm orb}$ the orbital mean motion. 

Equations~(\ref{tidal_potential}) and~(\ref{tidal_potential_comp}) give the expression of the tidal potential due to the planet inside the star ($r\leq R$) in the most general case of an eccentric and oblique orbit. If the orbit is circular and co-planar ($e=0$, $i=0$), the only nonvanishing terms have $n=m$ and $l-m$ must be even. If the orbit is circular $(e=0$), but oblique ($i \not=0$), then $n$ ranges between $-l$ and $l$, while $l-n$ must be even. In the case of an eccentric orbit, all the Fourier components of the series development of the potential, each corresponding to a different $n$, are required \citep[for details see][]{Ogilvie13,Ogilvie14}. However, the coefficients $A_{lmn}$ decrease with increasing order $n$, and therefore only the terms with $|n| \leq (l+p)$ are required if components smaller than $O(e^{p})$ can be neglected, where $p$ is any positive integer \citep{MathisLePoncin09,Ogilvie14}. 

Each component of the tidal potential in the frame rotating with the star has a frequency $\omega_{mn} \equiv m\Omega_{\rm s} - n\Omega_{0}$ that is an integer linear combination of the spin $\Omega_{\rm s}$ and orbital $\Omega_{0}$ frequencies. As we shall see in Sect.~\ref{forced_osc}, the resonant excitation of forced oscillations in the stellar magnetic field occurs when the frequency of one of the tidal components coincides with a characteristic frequency of the magnetohydrodynamic (MHD) system that is much smaller than the spin frequency. In other words, we find that a resonant excitation of the magnetic field oscillations is possible when $ | \omega_{mn} | \ll \Omega_{\rm s}$, which corresponds to a rational ratio $n/m$ between the two frequencies $\Omega_{\rm s}$ and $\Omega_{0}$, thus accounting for the spin-orbit commensurability observed in those systems. 

The values of $m$ and $n$ for the resonant frequency are not small because we invoke resonances with $m$ up to $8$, which implies that $l$ must be at least equal or greater. Therefore, the corresponding tidal component will be much smaller than the quadrupolar component with $l=2$ that dominates the tidal interaction \citep{Ogilvie14}. 

For the sake of simplicity, we focus on the case of a point-like planet on a circular oblique orbit. In general, the dissipation of the tides raised by the star inside the planet is capable of damping the orbit eccentricity on a timescale short in comparison with the main-sequence lifetime of the host star. On the other hand, the damping of the obliquity is due to the tidal dissipation inside the star that requires remarkably longer timescales \citep[e.g.,][]{Leconteetal10,Ogilvie14}. Therefore, our simplifying assumption of a circular oblique orbit is justified in most of the systems. In this case,
\begin{equation}
A_{lmn} (0, i) = (-1)^{n} \frac{4\pi}{2l+1} Y_{l}^{-n} \left( \frac{\pi}{2}, 0 \right) \, d_{m n}^{l} (i),  
\label{amnl_eq}
\end{equation}
where $d_{m n}^{l} (i)$ are the elements of the Wigner's $\bf d$ matrix introduced in Appendix~\ref{potential_development}, where a full derivation of Eq.~(\ref{amnl_eq}) is given. 

\begin{figure}
%\hspace*{-7mm}
 \centering{
 \includegraphics[width=5cm,height=7cm,angle=90,trim=55 97 70 48,clip]{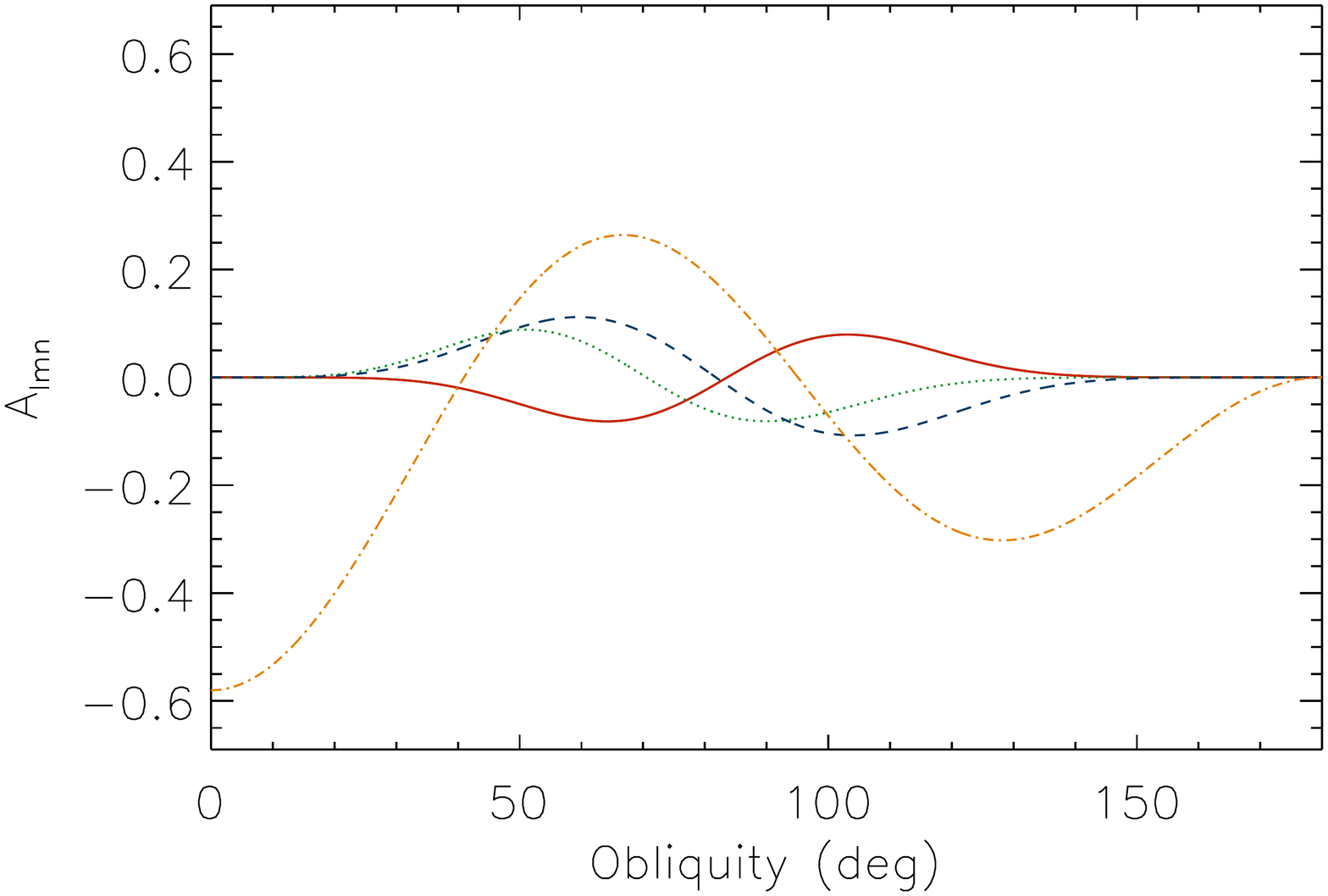}} % almn_corr.pdf}} % HD106252_1_rhkcorr_KR_final.pdf}} [trim=87 87 93 93, clip] trim=20 85 50 60,clip  20 87 60 48
% \vspace*{-10mm}
   \caption{Coefficients $A_{lmn}$ of the series development of the tidal potential vs. the obliquity of the circular planetary orbit for $l=9$, $m=8$, and $n=1$ (solid red line), corresponding to the case of Kepler-17; $l=9$, $m=8$, and $n=3$, corresponding to CoRoT-2 (dotted green line); $l=7$, $m=6$, and $n=1$, corresponding to HAT-P-11 (dashed blue line); and $l=3$, $m=1$, and $n=1$, corresponding to the synchronous systems CoRoT-4 and $\tau$ Boo (dash-dotted orange line). }
              \label{almn_plot}%
\end{figure}

In Fig.~\ref{almn_plot} we plot $A_{lmn}$ versus the obliquity $i$ of the system for some representative cases taken from Table~\ref{table1}. For each of the cases, the values of $m$ and $n$ are fixed by the observed $P_{\rm rot}/P_{\rm orb}$ ratio and the minimum allowed $l$ is assumed. It is equal to $m$ or $m+1$ because $l-n$ must be even, assuming a circular orbit. 
%%%%%%%%%%%%%%%%%%%%%%%%%%%%%%%%%%%%%
%%%%%%%%%%%%%%%%%

Considering a Sun-like star and assuming that the magnetic field is stored in the overshoot layer at the base of the convection zone, $r/R \sim 0.7$, which implies a ratio of the component with $l=8$ to that with $l=2$ at least on the order of $(0.7)^6 \sim 0.12$, assuming that the $A_{lmn}$ coefficients have comparable values. Since the quadrupolar potential ($l=2$) is already very small in the case of a close-by star-planet system, being on the order of $\left( m_{\rm p}/M \right) \left(R/a \right)^{3} \la 10^{-6}$ of the stellar gravitational potential, one may wonder whether a component of the stellar potential smaller than $10^{-9}-10^{-7}$  can excite resonant oscillations of the interior magnetic field. %\LEt{ Verify that your intended meaning has not been changed.}
We assumed that this is possible thanks to the extremely small dissipation inside the overshoot layer of the star that allows the small periodic forcing to accumulate its effects for a sufficiently long time interval to produce sizable oscillations of the magnetic field, which we describe in detail in the next sections. 

\subsection{Internal magnetic field of the star}
\label{internal_m_field}

Interior magnetic fields in stars are subject to magnetic buoyancy and tend to emerge when they are located inside the convection zones because of their superadiabatic stratification. Conversely, they can be stored for long time intervals into the radiative zone and in the overshoot layer below the outer convection zone of the Sun or solar-like stars thanks to their subadiabatic stratification. In those layers, the emersion is possible on the very long thermal timescale because thermal conduction must equilibrate their temperature with that of the surrounding medium before magnetic buoyancy can become effective \citep[][]{SchmittRosner83,Schmittetal84}. Therefore, the overshoot layer is regarded as a possible location for the solar or stellar hydromagnetic dynamos \citep[e.g.,][]{Parker93}. 

The subadiabatic stratification and the physical state of the overshoot layer are governed by the penetration of the convective fluid motions coming from the above convection zone that do not stop at the level where the gradient $\nabla$ becomes equal to the adiabatic gradient $\nabla_{\rm a}$, but continue to penetrate in the underlying stably stratified layers  because of their inertia. Unfortunately, a detailed description of undershooting convection in the Sun and solar-like stars is very difficult and is made even more complicated by the strong radial shear in the layer discovered by helioseismology, the likely intense magnetic fields stored there, and possible molecular weight gradients \citep[e.g.,][]{Zahn91,SpiegelZahn92}.  Therefore, we must make assumptions on the values of the subadiabatic gradient $\delta \equiv \nabla -\nabla_{\rm a} < 0$ and on the coefficients expressing the viscosity $\nu$, the magnetic diffusivity $\eta$, and the thermal diffusivity $\kappa$ in the layer. 

Our fundamental assumption is that the overshoot layer below  the stellar convection zone is the site of a magnetized layer whose field evolves very slowly, that is, on a timescale much longer than the inverse of the tidal frequency $\hat{\omega}_{nm} \equiv n\Omega_{0} -m\Omega_{\rm s}$ responsible for the resonant excitation of the field oscillations.  

A model of the response of such a magnetized layer to the time-varying tidal potential is made very complex by the gradient of the field intensity, the density stratification, and the differential rotation of the star. A useful simplification is to treat the layer as an ensemble of isolated slender magnetic flux tubes and focus on one of them that will be in resonance with the tidal potential.  Specifically, we assume an initial equilibrium state with an isolated axisymmetric slender magnetic flux tube having a purely azimuthal field ${\vec B} = B {\vec e}_{\phi}$  located in the equatorial plane at a radial distance $r_{0}$ from the barycenter of the star, where ${\vec e}_{\phi}$ is a unit vector in the azimuthal direction (see Fig.~\ref{flux_tube_pict}).  The flux tube is slender in the sense that its diameter is much smaller than its curvature radius (coincident with $r_{0}$) and the local pressure scale height, $H$, which makes the equations governing its evolution under the effect of an external perturbation particularly simple. 

The viscosity and the magnetic diffusivity in the overshoot layer are poorly known because of our limited understanding of the physical processes occurring inside it. In this work, we adopt the values in Table~1 of \citet{BrunZahn06}, that is, $\nu \sim 3\times 10^{-3}$ m$^{2}$~s$^{-1}$ and $\eta \sim 0.1$ m$^{2}$~s$^{-1}$. A mild turbulence in the layer could accelerate diffusion, but we assume that the magnetic field is capable of damping turbulence inside the flux tube, so molecular plasma values are adequate for the diffusivities. The timescale for the diffusion, $t_{\rm diff}$, of the magnetic field of a slender flux tube of diameter $d_{\rm ft} \ga 5 \times 10^{6}$~m is $t_{\rm diff} \sim d_{\rm ft}^{2}/\eta \ga 3 \times 10^{14}$~s, that is, $\ga 10^{7}$ yr. Such a long lifetime plays a crucial role in our model because it allows the oscillations excited by the extremely small tidal potential to be slowly amplified at resonance until they reach a significant amplitude. 

The thermal timescale of the flux tube is on the order of $d_{\rm ft}^{2}/\kappa \sim 10^{10}$~s, where $\kappa \sim 10^{3}$~m$^{2}$~s$^{-1}$ is the thermal diffusivity. Such a timescale is shorter than the magnetic diffusion timescale, {implying that our  flux tube must be considered in thermal equilibrium with the surrounding medium. In this case, the flux tube experiences a radial buoyancy force that tends to move it upward outside of the overshoot layer into the convection zone, where it becomes dynamically unstable. We can assume that the flux tube is held in a stable equilibrium inside the overshoot layer  thanks to the action of the downward directed convective plumes that penetrate into the stable layer from the overlying convection zone. Such a convective pumping has been shown to be an effective mechanism to keep magnetic flux tubes inside the overshoot layer provided that magnetic buoyancy is smaller than the hydrodynamic force produced by the plume downdrafts that can be approximated as an aerodynamic drag \citep[see, e.g., Sect.~6 of][and references therein]{Fan21}. Specifically, a flux tube with a magnetic field intensity $B$ can be held inside the overshoot layer if
\begin{equation}
\frac{B^{2}}{2\mu H} \la C_{\rm D} \frac{\rho_{0} v_{\rm d}^{2}}{\pi d_{\rm ft}},
\label{eq_downdraft}
\end{equation}
where $\mu$ is the magnetic permeability of the plasma, $\rho_{0}$ its density, $v_{\rm d}$ the velocity of the downdrafts at the level $r_{0}$ where the flux tube is located, and $C_{\rm D} \sim 1$ the aerodynamic drag coefficient. In a Sun-like star, typical values are $H \sim 6 \times 10^{7}$~m, $\rho_{0} \sim 200$~kg~m$^{-3}$, while $v_{\rm d}$ decreases from $\sim 100$ to $ \sim 1$~m~s$^{-1}$ from the top to the bottom of the overshoot layer \citep[cf.][]{Zahn91,FerrizMasSchussler93}. Considering a diameter $d_{\rm ft} \sim 5 \times 10^{6}$~m for the flux tube, it can be held stable inside the lower part of the overshoot layer if its magnetic field is smaller than $\sim 1.6 \times 10^{3}$~G, while fields up to a few $10^{4}$~G require a smaller diameter or can be stored close to the boundary with the overlying convection zone. In the slender flux tube approximation, $d_{\rm ft} \ll H_{\rm p}$; therefore, the equilibrium magnetic field can be stronger than the equipartition field, $B_{\rm eq} \sim \sqrt{\mu \rho_{0}}\,  v_{\rm d}$, ensuring that the downdrafts do not have enough kinetic energy to destroy the flux tube. }%but we shall assume that the buoyancy in the equatorial plane is equilibrated by the magnetic tension or the centrifugal forces acting on the flux tube \citep[for a general discussion of the stability under the effects of those forces see][]{MorenoInsertisetal92}. 

We do not discuss the physical processes that can produce slender magnetic flux tubes in the overshoot layer and refer the reader to the reviews of \citet{Schussler05} and \citet{Fan21} for more information. We only assumed that a toroidal slender flux tube is located there because the subadiabatic stratification of the layer {and the convective downdrafts} can hold it stable against magnetic buoyancy, at least if its magnetic field intensity is below the threshold for the development of an undulatory instability, which is $\sim 10^{5}$~G in the Sun or  higher fields in more rapidly rotating stars \citep{FerrizMasSchussler94,Schussleretal94,Caligarietal95,Granzeretal00}. As we shall see, the flux tubes where resonant oscillations can be excited have a magnetic field strength $\la 10^{3}-10^{5}$~G; thus, their stability to undulatory  perturbations is in general not an issue (cf. Sect.~\ref{applications}). 

Owing to their very slow evolution and the lack of instability leading to their emergence, the resonant magnetic flux tubes considered in the present model cannot contribute to stellar magnetic cycles with typical periodicities of $\sim 10-20$ years. We speculate that such cycles, and even shorter ones as suggested by observations of Kepler-17, can be produced by a dynamo operating in the convection zone or in another domain of the overshoot layer, possibly located above the layer where our stable flux tubes are stored. The possibility of different dynamos operating in different stellar layers (and with different cycle periods) has been suggested by, for example, \citet{Brandenburgetal17}. %\LEt{ We do not allow the use of "e.g." or "i.e." within the main text (in parentheses or within figure/table captions is fine).}

The need for having different flux tubes that produce starspots in the stellar photospheres or that can resonate with the time-varying tidal potential is dictated by the short dynamical timescale of the starspots, which become unstable and emerge from the overshoot layer on a timescale shorter than the stellar cycles, typically on the order of one year \citep{Caligarietal95,Fan21}. %\LEt{ "Former" and "latter" should ideally only be used to distinguish between two items in a clear set of two, otherwise there is room for ambiguity and/or confusion. Verify that your intended meaning has not been changed.}
We can speculate that the flux tubes that can resonate with the tidal potential are the low-field tail of the distribution of slender flux tubes produced by the stellar dynamo (with $B \la 10^{3}-10^{4}$~G); these tubes can be more easily pumped down to the lower part of the overshoot layer, where they can reside for a very long time, up to $\sim 10^{7}$ years, thus giving the resonant oscillations the possibility to grow to a sizeable amplitude (cf. Sect.~\ref{amplification}). %\LEt{ Verify that your intended meaning has not been changed.} 
They do not become unstable and do not emerge to the photosphere, but remain confined into the overshoot layer even when they reach their maximum oscillation amplitude. However, their oscillations can perturb the flux tubes with a stronger field, triggering their instability; this will make them emerge in the photosphere (see also Sect.~\ref{secondary_reg_mechanisms}). 

\subsection{Equilibrium state and linearized equations of motion of a slender magnetic flux tube}

The equilibrium state of our toroidal slender flux tube in the equatorial plane of our star has been described by \citet{FerrizMasSchussler93} and \citet{FerrizMasSchussler94}. They introduce the equations governing the flux tube evolution that are particularly simple because the tube is described as a bundle of magnetic field lines separated by the surrounding unmagnetized medium by a surface across which the field goes discontinuously to zero in the ideal (i.e., dissipationless) MHD approximation. 

The forces acting on the flux tube are the external pressure gradient, the gradient of the gravitational potential, and the magnetic tension force due to the curvature of the field lines together with the Coriolis force and the centrifugal force because our reference frame is rotating with the frequency  $\Omega_{\rm s}$ with respect to an inertial reference frame. We chose $\Omega_{\rm s}$ to be the angular velocity of the matter inside the unperturbed flux tube. Since the star can rotate differentially with an angular velocity $\Omega_{\rm e} (r)$, the external velocity field in the equatorial plane is $ r [\Omega_{\rm e} (r) - \Omega_{\rm s}]{\vec e}_{\phi}$. 

\begin{figure}
%\hspace*{-7mm}
 \centering{
 \includegraphics[width=9cm,height=6cm,angle=0,trim=85 85 85 85,clip]{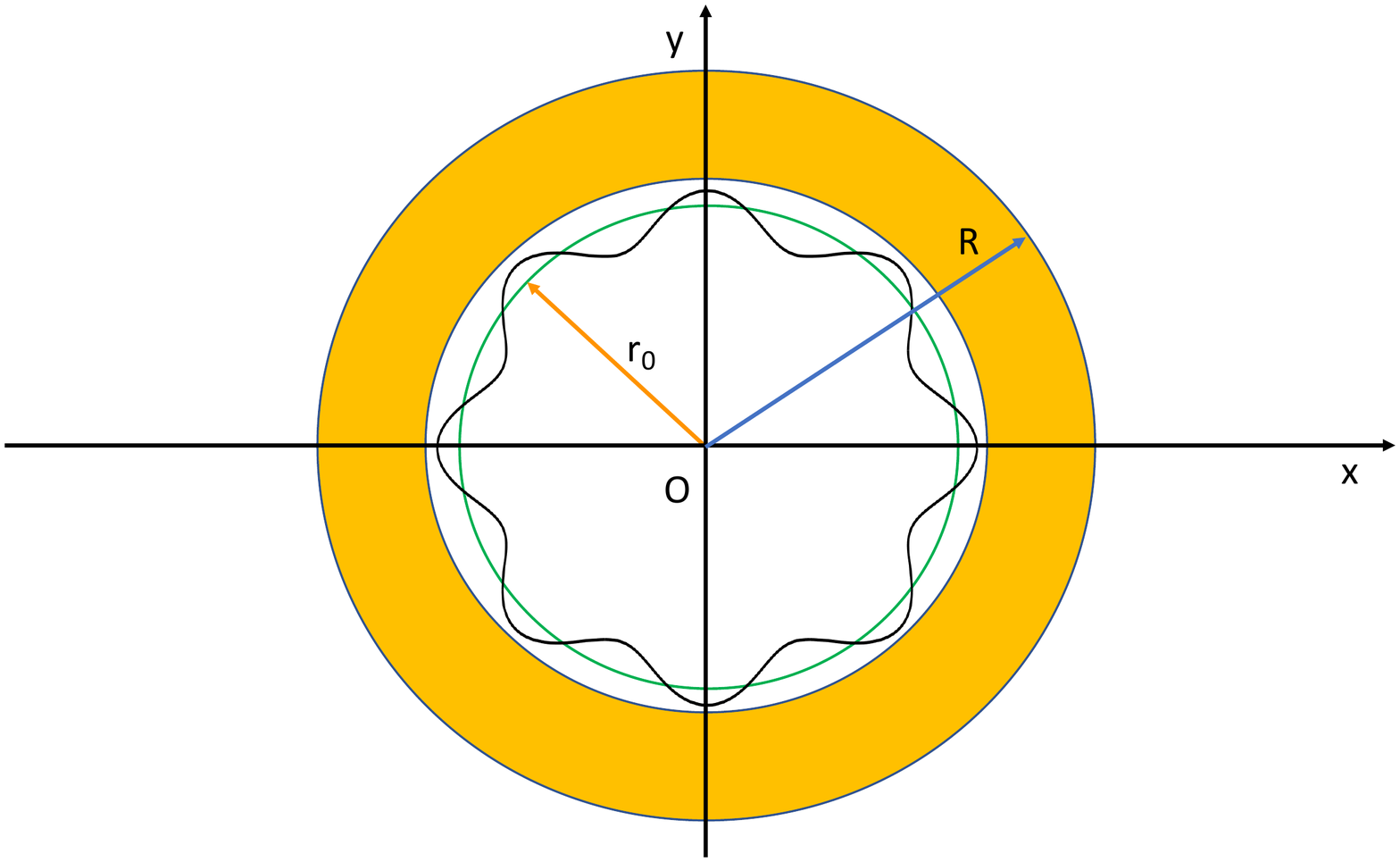}} % flux_tube_picture_3.pdf}} % HD106252_1_rhkcorr_KR_final.pdf}} [trim=87 87 93 93, clip]
% \vspace*{-10mm}
   \caption{Sketch of a slender magnetic toroidal flux tube in equilibrium in the equatorial plane of a solar-like star (solid green line).  The $xy$ plane is the equatorial plane of the star as seen from the top of the vertical $z$ axis. The barycenter of the star is $O$ and its radius is $R$, while $r_{0}$ is the radius of the unperturbed flux tube. The convection zone of the star is rendered in orange, while the radiative zone, including the overshoot layer where the flux tube is located, is rendered in white. The flux tube perturbed by an  oscillation mode with $m=8$ is shown as a solid black line.  } 
\label{flux_tube_pict}%
\end{figure}

The axisymmetric equilibrium state of the flux tube in an isolated star is described by the equation (cf. Fig.~\ref{flux_tube_pict})
\begin{equation}
\frac{v_{\rm A}^{2}}{r_{0} g_{0}} - \left( \frac{\rho_{\rm e0}}{\rho_{\rm i0}}- 1\ \right) \left(1 - \frac{r_{0} \Omega_{\rm e0}^{2}}{g_{0}} \right) + \frac{r_{0}}{g_{0}} \left( \Omega_{\rm e0}^{2} - \Omega_{\rm s}^{2} \right) = 0,  
\label{equilibrium}
\end{equation}
 where $v_{\rm A} \equiv B/\sqrt{\mu \rho_{\rm i0}}$ is the Alfven velocity inside the flux tube, $\mu$ the magnetic permeability of the plasma, $r_{0}$ the radius of the flux tube, $g_{0}$ the acceleration of gravity at the distance $r_{0}$ from the center of the star, $\rho_{\rm e0}$ the unperturbed density of the surrounding medium, $\rho_{\rm i0}$ the internal unperturbed density of the flux tube, $\Omega_{\rm e0} =\Omega_{\rm e}(r_{0})$, and in general the subscript $0$ indicates unperturbed quantities at the distance $r_{0}$ from the center of the star. Equation~(\ref{equilibrium}) expresses the balance between the curvature force, that is, the magnetic tension expressed in terms of the Alfven velocity, and the buoyancy and centrifugal forces in the equilibrium state of the flux tube \citep[see][for details]{MorenoInsertisetal92,FerrizMasSchussler93}. 
%%%%%%%%%%%%%%%%%%%%%%%%%%%%%%%%%%%%%
%%%%%%%%%%%%%%%%%
 
 The  small perturbations around the equilibrium state have been investigated by \citet{FerrizMasSchussler93}. % who found that the equations describing perturbations within the equatorial plane and perpendicular to the equator are decoupled from each other. This allows us to decouple the study of the perturbations due to the tidal potential considering the motions in the equatorial plane independently from those in the latitudinal direction. 
 We generalize their nondimensional equations of motion  for a free magnetic flux tube to the case of a flux tube subject to an external tidal force $-\rho_{0 i}\nabla \Psi$  per unit volume and a small viscous damping as
\begin{eqnarray}
\tau^{2} \left( \ddot{\xi}_{\phi} + 2\Omega_{\rm s} \dot{\xi}_{r} + \frac{1}{r_{0}}\frac{\partial \Psi}{\partial \phi_{0}} + 2 \zeta \dot{\xi}_{\phi} \right)  & = & \nonumber \label{eq_motion1} \\ 
2 \frac{c_{\rm T}^{2}}{v_{\rm A}^{2}} f^{2} \frac{\partial^{2} \xi_{\phi}}{\partial \phi_{0}^{2}} + 4f \frac{c_{\rm T}^{2}}{v_{\rm A}^{2}} \left( f - \frac{1-x}{2\gamma} \right) \frac{\partial \xi_{r}}{\partial \phi_{0}},  \\ 
\tau^{2} \left( \ddot{\xi}_{r} - 2\Omega_{\rm s} \dot{\xi}_{\phi} + \frac{\partial \Psi}{\partial r} + 2 \zeta \dot{\xi}_{r} \right)  & = & \nonumber \\ 
2 f^{2} \frac{\partial^{2} \xi_{r}}{\partial \phi_{0}^{2}} -4f \frac{c_{\rm T}^{2}}{v_{\rm A}^{2}} \left( f - \frac{1-x}{2\gamma} \right) \frac{\partial \xi_{\phi}}{\partial \phi_{0}} + T \xi_{r}, 
\label{eq_motion2} \\
\tau^{2} \left( \ddot{\xi}_{\theta} + \frac{1}{r_{0}}\frac{\partial \Psi}{\partial \theta} +2 \zeta \dot{\xi_{\theta}} \right) = 2 f^{2} \frac{\partial \xi_{\theta}}{\partial \phi_{0}^{2}} + \frac{2 f^{2}+ f\beta(x_{\rm e} -x)}{1-x_{\rm e}} \xi_{\theta}
\label{eq_motion3}
,\end{eqnarray}
where a dot indicates the time derivative; $\vec \xi = (\xi_{\phi}, \xi_{r}, \xi_{\theta})$ is the small displacement with respect to the position of equilibrium; $\phi_{0}$   is the azimuthal coordinate measured along the unperturbed flux tube;  $\tau$ is a characteristic timescale defined as 
\begin{equation}
\tau \equiv \left( \frac{\beta H}{g_{0}} \right)^{1/2},
\label{tau_defin}
\end{equation}
with $\beta \equiv 2\mu p_{\rm i0}/B^{2}$, $p_{\rm i0}$ being the unperturbed internal pressure in the flux tube; $c_{\rm T}$ is the tube wave speed
\begin{equation}
c_{\rm T} \equiv \frac{c_{\rm s} v_{\rm A}}{\sqrt{c_{\rm s}^{2} + v_{\rm A}^{2}}},
\end{equation}
with $c_{\rm s} = \sqrt{\gamma p_{\rm i0}/\rho_{\rm i0}}$ being the internal sound speed and $\gamma = 5/3$ the adiabatic exponent; 
$\zeta$ is the inverse of the viscous damping timescale; $f\equiv H/r_{0}$; $x \equiv  r_{0} \Omega_{\rm s}^{2}/g_{0}$; $x_{\rm e} \equiv r_{0} \Omega_{\rm e0}^{2}/g_{0}$; and $T$ is given in the limit $\beta \gg 1$ by
\begin{equation}
T \equiv 2 (\sigma -1) f^{2} + \frac{1}{\gamma} \left( 4f - \frac{2}{\gamma} +1 \right) + \beta \delta  - (1-\sigma) \tau^{2} \left( \Omega_{\rm e0}^{2} - \Omega_{\rm s}^{2} \right)  
- 4q \tau^{2} \Omega_{\rm e0}^{2},  
\label{T_eq}
\end{equation}
where $ \sigma \equiv \left[ d \log g(r) / d \log r \right]_{r_{0}} $ expresses the dependence of the acceleration of gravity on the radial distance and $q \equiv r_{0} \Omega^{\prime}_{\rm e}(r_{0})/[2 \Omega_{\rm e}(r_{0})]$ measures the shear due to the radial differential rotation at $r=r_{0}$ with the prime indicating the derivative with respect to the radial coordinate. 

We see that Eqs. (\ref{eq_motion1}) and (\ref{eq_motion2}), which rule the motion in the equatorial plane, are decoupled from Eq.~(\ref{eq_motion3}), which governs the motion in the direction of colatitude.\ Therefore, they can be solved independently from each other.  To solve Eqs.~(\ref{eq_motion1}) and~(\ref{eq_motion2}), we considered an individual mode of oscillation of the form
\begin{equation}
{\vec \xi} = \hat{\vec \xi} \exp (j\omega t + jm\phi_{0}), 
\label{xieq}
\end{equation}
where $\hat{\vec \xi}$ is a constant vector, $t$ the time, and the periodicity of the solution in the azimuthal direction imposes that $m$ be an integer (see Fig.~\ref{flux_tube_pict}).
Before treating the case of forced oscillations excited by the time varying tidal potential, it is useful to study the conservation of energy and the free modes of oscillations of the system governed by Eqs. (\ref{eq_motion1}) and (\ref{eq_motion2}). 

\subsection{Mechanical energy conservation}
We can derive an expression for the variation in the nondimensional total mechanical energy for the motions restricted to the equatorial plane by multiplying Eq.~(\ref{eq_motion1}) by $\dot{\xi}_{\phi}$, Eq.~(\ref{eq_motion2}) by $\dot{\xi}_{r}$, and summing them up. We neglect the viscous dissipation terms proportional to $\zeta$ because they are extremely small, as we shall see in Sect.~\ref{forced_osc}, and consider a displacement given by Eq.~(\ref{xieq}) so that $\partial /\partial t \rightarrow j\omega$ and $\partial/\partial \phi_{0} \rightarrow jm$. In this way, we obtain
%\begin{equation}
%\frac{d{\cal E}}{dt} = 4 f \frac{c_{\rm T}^{2}}{v_{\rm A}^{2}} \left( f - \frac{1-x}{2\gamma} \right) \left( \dot{\xi}_{\phi} \frac{\partial \xi_{r}}{\partial\phi_{0}} - \dot{\xi}_{r} \frac{\partial \xi_{\phi}}{\partial\phi_{0}} \right) - \tau^{2} \dot{\vec \xi} \cdot \nabla \Psi
%\label{energy_equation}
%\end{equation}
\begin{equation}
\frac{d{\cal E}}{dt} =  - \tau^{2} \dot{\vec \xi} \cdot \nabla \Psi
\label{energy_equation}
,\end{equation}
where the total mechanical energy density is defined as
\begin{eqnarray}
{\cal E} = \frac{1}{2} \tau^{2} \left( \dot{\xi}_{\phi}^{2} + \dot{\xi}_{r}^{2} \right) + f^{2} m^{2} \left( \frac{c_{\rm T}^{2}}{v_{\rm A}^{2}} \xi_{\phi}^{2} + \xi_{r}^{2} \right) - \frac{1}{2} T \xi_{r}^{2}.  
%{\cal E} = \frac{1}{2} \tau^{2} \left( \dot{\xi}_{\phi}^{2} + \dot{\xi}_{r}^{2} \right) - f^{2} \frac{\partial^{2}}{\partial \phi_{0}^{2}} \left( \frac{c_{\rm T}^{2}}{v_{\rm A}^{2}} \xi_{\phi}^{2} + \xi_{r}^{2} \right) - \frac{1}{2} T \xi_{r}^{2}.  
\label{tot_energy_mode}
\end{eqnarray}
The dimensional form of the energy density can be obtained from Eq.~(\ref{tot_energy_mode}) by multiplying both sides by $\rho \tau^{-2}$ and introducing a dimensional displacement vector $\vec \xi$. 
The energy density consists of the sum of the kinetic energy, the potential energy due to the magnetic tension force -- the term proportional to $ f^{2} m^{2}$ times the squares of the components of the displacement -- and the energy associated with the stratification and the centrifugal effects as expressed by the term proportional to $T$. The right hand side of Eq.~(\ref{energy_equation}) represents the work done by the tidal force per unit volume and time. 

\subsection{Free oscillation modes and stability of the magnetic flux tube}
The free oscillation modes are the solutions of Eqs. (\ref{eq_motion1}) and (\ref{eq_motion2}) when the external forcing is absent, that is, $\nabla \Psi = 0$. For simplicity, we considered the ideal case without dissipation ($\zeta =0$) and studied the stability of the system in the case of small motions confined to the equatorial plane, summarizing the results already obtained by \citet{FerrizMasSchussler93} and \citet{FerrizMasSchussler94}. In the case of free oscillations, the frequency $\omega$ is in general a complex number, the imaginary part of which determines the stability of a normal mode with a given azimuthal wavenumber, $m$. When the imaginary part of $\omega$ is negative, the mode amplitude grows in time and the mode is unstable, while it is stable when it is positive. 

The homogeneous system of  partial differential Eqs. (\ref{eq_motion1}) and (\ref{eq_motion2}) can be reduced to an homogeneous system of algebraic equations by substituting Eq.~(\ref{xieq}) and taking into account that $\partial/\partial t \rightarrow j\omega $, while $\partial/\partial \phi_{0} \rightarrow jm $. The frequency $\omega$ comes from the vanishing of the determinant of this system of algebraic equations. This condition can be written as a fourth-order  polynomial equation (the so-called characteristic equation), 
\begin{equation}
P(\tilde{\omega}) \equiv  \tilde{\omega}^{4} + d_{2} \tilde{\omega}^{2} + d_{1} \tilde{\omega} + d_{0} =0,
\label{char_eq}
\end{equation}
where $\tilde{\omega} \equiv \tau \omega$ and 
\begin{eqnarray}
d_{2} = 2 f^{2} (\sigma -1 - 2m^{2}) + \frac{4}{\gamma} f - \frac{2}{\gamma} \left( \frac{1}{\gamma} - \frac{1}{2} \right) + \beta \delta \nonumber \\ 
+ (\sigma -1) \left( \tilde{\Omega}_{\rm e0}^{2} - \tilde{\Omega}_{\rm s}^{2} \right) - 4 \tilde{\Omega}_{\rm s}^{2} - 4q \tilde{\Omega}_{\rm e0}^{2}, \\
d_{1} = 16 m f \left( f - \frac{1}{2\gamma} \right) \tilde{\Omega}, \\
d_{0} = -2m^{2} f^{2} \left[ d_{2} + 2 m^{2} f^{2} + 8 \left(f - \frac{1}{2\gamma} \right)^{2} + 4 \tilde{\Omega}_{\rm s}^{2} \right] , 
\end{eqnarray}
where $\tilde{\Omega}_{\rm e0} = \tau \Omega_{\rm e0}$ and $ \tilde{\Omega}_{\rm s} = \tau \Omega_{\rm s}$. The necessary and sufficient condition for stability can be written as
\begin{equation}
- \frac{4}{27} d_{0} \left( d_{2}^{2} - 4d_{0} \right)^{2} + d_{1}^{2} \left( \frac{d_{2}^{3}}{27} - \frac{4}{3} d_{0} d_{2} + \frac{d_{1}^{2}}{4} \right) < 0 
\label{ns_stability_condition}
\end{equation}
\citep[see][]{FerrizMasSchussler93,FerrizMasSchussler94}.

\subsection{Forced oscillations of the magnetic flux tube}
\label{forced_osc}

The forcing produced by the mode of the tidal potential with frequency $\omega_{mn}$ can be written as 
\begin{eqnarray}
\tau^{2} \frac{1}{r_{0}} \frac{\partial \Psi_{lmn}}{\partial \phi_{0}} = \hat{p}_{\phi} \exp (j\omega_{mn} t + jm \phi_{0}) \label{forcing1} \\
\tau^{2} \frac{\partial \Psi_{lmn}}{\partial r} = \hat{p}_{r} \exp (j\omega_{mn} t + jm \phi_{0}), 
\label{forcing}
\end{eqnarray}
where the complex quantities $(\hat{p}_{\phi}, \hat{p}_{r})$ can be deduced from the expression of the potential in Eq.~(\ref{tidal_potential_comp}), while the tidal frequency $\omega_{mn} = m \Omega_{\rm s} - n \Omega_{0}$ is real.  We considered only the stationary solution of the system of equations, assuming that the effects of the initial conditions have had time to decay under the action of the viscous dissipation. Alternatively,  the same solution is obtained if the oscillations are excited from an initial state of vanishing amplitude. % forcing slowly increases the amplitude of the oscillations starting from a state of vanishing initial velocity. 

Substituting Eqs.~(\ref{xieq}), (\ref{forcing1}), and (\ref{forcing}) into Eqs. (\ref{eq_motion1}) and (\ref{eq_motion2}), it is possible to reduce the latter two to an algebraic system with $\omega = \omega_{mn}$ and the same $m$ of the component of the tidal potential,
\begin{equation}
\left( \begin{array}{cc}
A - 2j\tilde{\omega}_{mn} \tilde{\zeta}  &  -jD \\
jD & C - 2j \tilde{\omega}_{mn} \tilde{\zeta} \\
\end{array} \right)
 \left( \begin{array}{c} \hat{\xi}_{\phi} \\ 
\hat{\xi}_{r} 
\end{array}
\right) = 
\left( \begin{array}{c} \hat{p}_{\phi} \\ 
\hat{p}_{r} 
\end{array}
\right),
\label{forced_osc_alg}
\end{equation}
where $\tilde{\omega}_{mn} = \tau \omega_{mn}$, $\tilde{\zeta} = \tau \zeta$,  
\begin{eqnarray}
A \equiv \tilde{\omega}_{mn}^{2} - 2 f^{2} m^{2}, \\
D \equiv  2 \left [\tilde{\Omega}_{\rm s} \tilde{\omega}_{mn} - 2 mf \left( f - \frac{1-x}{2 \gamma} \right) \right] ,  \mbox{and} \\
C \equiv A + T.  
\end{eqnarray}
The solution of the algebraic system (\ref{forced_osc_alg}) that gives the amplitude of the stationary forced oscillations is
\begin{eqnarray}
\hat{\xi}_{\phi} = W \left\{ C P(\tilde{\omega}_{mn}) \hat{p}_{\phi} - 2DE \tilde{\omega}_{mn} \tilde{\zeta} \hat{p}_{r} + \right. \nonumber  \label{xi_phi}\\
\left. j \left[ DP(\tilde{\omega}_{mn}) \hat{p}_{r} +2\tilde{\omega}_{mn} \tilde{\zeta} \left(CE- P(\tilde{\omega}_{mn}) \right)  \hat{p}_{\phi}  \right] \right\}, \\ 
\hat{\xi}_{r} = W \left\{ A P(\tilde{\omega}_{mn}) \hat{p}_{r} + 2DE \tilde{\omega}_{mn} \tilde{\zeta} \hat{p}_{\phi} + \right. \nonumber  \label{xi_r}\\
\left. - j \left[ DP(\tilde{\omega}_{mn}) \hat{p}_{\phi} -2\tilde{\omega}_{mn} \tilde{\zeta} \left(AE- P(\tilde{\omega}_{mn}) \right)  \hat{p}_{r}  \right] \right\}, 
\end{eqnarray}
where the amplification factor $W$ is 
\begin{equation}
W \equiv \frac{1}{[P(\tilde{\omega}_{mn})]^{2} + 4 E^{2} \tilde{\omega}_{mn}^{2} \tilde{\zeta}^{2}}, 
\label{w_eq}
\end{equation}
$E \equiv A + C = 2A+T$, 
and we considered only the first-order terms in $\tilde{\zeta}$ in Eqs.~(\ref{xi_phi}) and~(\ref{xi_r}), except in the expression of $W$. 

The amplitude of the forced oscillations is maximal when the real frequency $\tilde{\omega}_{mn}$ coincides with a zero of the polynomial $P(\tilde{\omega})$, which defines the condition of resonance. 
An important simplification is possible because we look for resonant oscillations of very low frequency that satisfies the condition $|\omega_{mn} | = | m\Omega_{\rm s} - n\Omega_{0}| \ll \Omega_{\rm s}$ in order to account for the close spin-orbit commensurability we want to explain with our model. This implies that we can neglect the term $\tilde{\omega}^{4}$ in the characteristic equation (\ref{char_eq}) making the calculations of its solutions straightforward. We considered the smaller of the two solutions in absolute value that is given by
\begin{equation}
\tilde{\omega}_{\rm res} = \frac{| d_{1} |-\sqrt{\Delta}}{2 | d_{2} |},
\label{magnetostrophic_sol}
\end{equation}
where the discriminant $\Delta \equiv d_{1}^{2} - 4 d_{2} d_{0} $ is required to be nonnegative for a real resonant frequency as in our case. 

We see that the frequency $\tilde{\omega}_{\rm res}$ is not only much smaller than $\tilde{\Omega}_{\rm s}$, but also of the nondimensional Alfven frequency $\tilde{\omega}_{\rm A} \equiv \sqrt{2} m f$. In other words, the resonant modes  we are considering can be analyzed in the framework of the so-called magnetostrophic approximation. It has been discussed in detail by \cite{FerrizMasSchussler94} in their Sect.~5.2 in the case of free oscillations. As we shall see in Sect.~\ref{applications}, in this regime the resulting magnetic field strength in the oscillating flux tube is on the order of $10^{2}-10^{4}$~G that corresponds to a high value of $\beta \ga 10^{5}-10^{10}$ and a long timescale $\tau \sim 10^{5} -10^{10}$~s. As a consequence, the dominating term in the mechanical energy density is that proportional to $T$ in Eq.~(\ref{tot_energy_mode}) and is on the order of $(1/2) (\beta |\delta|) \xi_{r}^{2}$ because $T \approx \beta |\delta |\sim 10^{2}-10^{5}$ for $|\delta |\ga  10^{-6}$. 
 
At resonance,  the stationary components of the displacement vector  are given by 
\begin{eqnarray}
\hat{\xi}_{\phi \, \rm res} =  \frac{1}{2 E \tilde{\omega}_{mn} \tilde{\zeta} } \left[ -D \hat{p}_{r} + j C \hat{p}_{\phi} \right], \\
\hat{\xi}_{r \, \rm res} =   \frac{1}{2 E \tilde{\omega}_{mn} \tilde{\zeta} } \left[ D \hat{p}_{\phi} + j A \hat{p}_{r} \right]. 
\end{eqnarray}
In the considered regime, the  term $T \approx \beta \delta$ dominates in the equations of the motion implying that $C \gg A, D$ and $ E \sim  C$. In this limit, the stationary solution at resonance becomes
\begin{eqnarray}
\hat{\xi}_{\phi \, \rm res} \simeq  \frac{j \hat{p}_{\phi}}{2  \tilde{\omega}_{mn} \tilde{\zeta} },  \nonumber \\
\hat{\xi}_{r \, \rm res} \simeq 0, 
\label{res_xi_components}
\end{eqnarray}
and the power per unit volume delivered by the time-varying tidal potential $\Psi_{lmn}$ in dimensional units is 
\begin{equation}
\left[ - \rho_{0} \dot{\vec \xi} \cdot \nabla \Psi_{lmn} \right]_{\rm res} \simeq \frac{\rho_{0}}{2 \zeta} \frac{\hat{p}_{\phi}^{2}}{\tau^{4}}  = \frac{\rho_{0}}{2 r_{0}^{2}\zeta} \left(\frac{\partial \Psi_{lmn}}{\partial \phi_{0}}\right)^{2}, %\frac{\tau}{2E \tilde{\zeta}} \left( C \hat{p}_{\phi}^{2} + A \hat{p}_{r}^{2} \right) \exp \left[ 2 \left( j\omega_{mn} t + 2jm \phi_{0} \right) \right].  
\label{res_power}
\end{equation}
where we neglected the density difference between the interior of the flux tube and the surrounding medium because it is very small for $\beta \gg 1$. 

The amplification factor $W$ of the forced oscillations in the stationary regime at resonance  is $(2\tilde{\omega}_{mn} \tilde{\zeta})^{-1}$ (cf. Eqs.~\ref{res_xi_components}) that is huge because $\tilde{\omega}_{mn}$ and $\tilde{\zeta}$ are very small. In particular, the timescale of viscous damping, $\zeta^{-1}$, is extremely long. It can be estimated by considering that the term expressing the viscous dissipation in the equation of Navier-Stokes is proportional to $\nu \nabla^{2} \dot{\vec \xi} $, where $\nu$ is the kinematic viscosity. Therefore, the timescale of viscous dissipation can be estimated as $t_{\rm diss} \sim 4 \pi^{2} r_{0}^{2}/(\nu m^{2})$ for the mode of azimuthal wavenumber $m$ that has a wavelength $2\pi r_{0}/m$. Considering a molecular value $\nu \sim 3 \times 10^{-3}$~m$^{2}$~s$^{-1}$, a Sun-like star ($r_{0} \sim 5 \times 10^{8}$~m), and $m=10$, we obtain $t_{\rm diss} \sim 10^{19}$~s, which can be neglected. Nevertheless, it is possible that a turbulent anisotropic viscosity be present in the overshoot layer outside of the flux tube with an horizontal component much larger than the vertical one \citep{SpiegelZahn92}, which justifies the consideration of the azimuthal length scale of the velocity field in the estimation of the damping time. Considering a turbulent value of $\nu_{\rm t} \sim 10^{4} \, \nu$ \citep{RuedigerKichatinov96,Denissenkovetal10}, the damping timescale $t_{\rm diss}$ becomes  $\sim 10^{15}$~s, which still implies a very large amplification factor. 

Specifically,  considering a magnetic field strength between $10^{2}$ and $10^{4}$~G, $\tau$ ranges between $10^{5}$ and $10^{10}$~s giving $\tilde{\zeta} \sim (\tau/t_{\rm diss}) \sim 10^{-10} -10^{-5}$, while $\tilde{\omega}_{mn} \sim 10^{-2}-10^{-3}$. In conclusion, the maximum amplification factor of the extremely small forcing produced by the tidal potential can vary over a wide range  on the order of $(2 \tilde{\omega}_{mn}\tilde{\zeta})^{-1} \sim 10^{7}-10^{13}$ even for a mode with $m=10$. However, this  amplification is obtained when the system reaches the stationary regime of  forced oscillations that may require a very long time interval. In the next subsection, we estimate such a time interval.  

 \subsection{Amplification of the resonant oscillations}
 \label{amplification}
We apply the conservation of mechanical energy to estimate the time required for the growth of the  forced oscillations investigated in the previous section. We assume that the system starts from a small oscillation amplitude, while keeping itself in resonance with the external tidal force for all the amplification interval (see Sect.~\ref{maintaining_osc} for details). In this way, the power per unit volume delivered by the tidal potential is maximal during each oscillation and the amplification timescale is the shortest possible. 

The ratio of the kinetic energy   to the magnetic tension energy in Eq.(\ref{tot_energy_mode}) is given by
\begin{equation}
\frac{\tau^{2} \dot{\xi}^{2}}{2 f^{2} m^{2} \xi^{2}} \sim \frac{\tilde{\omega}^{2}}{2 f^{2} m^{2}} \sim \frac{\tilde{\omega}^{2}}{\tilde{\omega}_{\rm A}^{2}} \ll 1
\end{equation}
in the low-frequency regime (cf. Sect.~\ref{forced_osc}). Therefore, we can neglect the kinetic energy of the oscillations in comparison to the magnetic tension and the stratification terms and write an approximate equation for the energy conservation so that
 the variation in the  mechanical energy of the resonant mode can be well approximated as 
\begin{equation}
\frac{1}{2} \left | \left(f^{2} m^{2} - T \right) \right| \frac{d\xi_{mn}^{2}}{dt} \sim \tau^{2}  \left| \dot{\vec \xi}_{mn} \cdot \nabla \Psi_{lmn} \right|
%2 m^{2} f^{2} \frac{d \xi_{mn}^{2}}{dt} = -\tau^{2} \dot{\vec \xi}_{mn} \cdot \nabla \Psi_{lmn}, 
\label{eq_amp}
,\end{equation}
where $\xi_{mn} $ is the mode displacement, $T$ has been defined in Eq.~(\ref{T_eq}), and we made the approximation that $\xi_{r} \sim \xi$. As a matter of fact, we saw that $|\xi_{r} | \ll | \xi_{\phi}| $ at resonance (cf. Eq.~\ref{res_xi_components}), but we introduce such an approximation to maximize the energy of the mode. 
From a physical point of view, the work done by the tidal force along the oscillations is used to store energy into the magnetic tension and overcome the stabilizing effect of the subadiabatic stratification in the layer where the flux tube is located.  

In the case of a circular oblique orbit, the module of $\nabla \Psi_{lmn}$ oscillates with a semiamplitude on the order of 
 \begin{equation}
|\nabla \Psi_{lmn} | \sim  m \frac{Gm_{\rm p}}{a^{2}} \left( \frac{r_{0}}{a}\right)^{l} \left| A_{lmn} (0, i) \, Y_{l}^{n}\left( \frac{\pi}{2} - i, 0 \right) \right|,  
\label{grad_psi}
\end{equation} 
where the coefficient $A_{lmn}$ is given by Eq.~(\ref{amnl_eq}) and the extrema of the gradient of the potential are reached when the planet is at the maximal angular distance from the equator of the star, that is, at a colatitude $ \frac{\pi}{2} \pm i$. In Eq.~(\ref{grad_psi}), we considered only the component of the gradient in the azimuthal direction because it is dominant at resonance in the considered regime (cf. Eq.~\ref{res_power}). 

Considering that the oscillations of $\dot{\vec \xi}$ and $-\nabla \Psi_{lmn}$ are in phase at resonance, $\xi_{r} \sim \xi$, and $\dot{\vec \xi}_{mn} = j\omega_{mn} {\vec \xi}_{mn}$, we can rewrite Eq.~(\ref{eq_amp}) in the approximate form
\begin{equation}
\dot{\xi}_{mn} \sim \frac{\tau^{2} \omega_{mn}}{|f^{2} m^{2} - T/2|} | \nabla \Psi_{lmn} |. % \sim \frac{H \omega}{g_{0} |\delta|} | \nabla \Psi_{lmn} |,
%\dot{\xi} \sim \frac{\tau^{2} \omega}{4 m^{2} f^{2}}| \nabla \Psi_{lmn} |. 
\end{equation}
This implies that an oscillation amplitude $\xi$ is attained at resonance after a time interval $\Delta t$ according to 
\begin{equation}
\xi_{mn} \sim  \frac{\tau^{2} \omega_{mn}}{|f^{2} m^{2} - T/2|} | \nabla \Psi_{lmn} |  \, \Delta t. %\frac{\tau^{2} \omega}{\beta |\delta|}| \nabla \Psi_{lmn} | \Delta t. 
\label{ampl_time}
\end{equation}
In Sect.~\ref{applications} we apply Eq.~(\ref{ampl_time}) to estimate the amplification timescale $\Delta t$ in the case of the systems listed in Table~\ref{table1} finding values on the order of $\Delta t \la 10^{6}-10^{7}$ yr for $\xi \sim 10^{6}$~m in the case of several systems. 

The above treatment does not include the effect of the turbulence produced in the surrounding medium as soon as the amplitude of the oscillations becomes sufficiently large. Specifically, the Reynolds number corresponding to an oscillation of amplitude $\xi$ is $Re \sim \omega \xi^{2}/\nu \sim 10^{6}$ for $\xi \sim 10^{6}$~m, $\omega \sim 10^{-9}$ s$^{-1}$, and $\nu \sim 3 \times 10^{-3}$ m$^{2}$~s$^{-1}$. Given that $Re \gg 1$, the motion of the flux tube produces turbulence in the surrounding unmagnetized plasma that can damp the oscillations.

In a stationary regime, the power provided by the tidal potential is compensated by the turbulent dissipation per unit volume that is given by ${\cal D} = 2 \rho \nu_{\rm t} (e_{kh})^{2}$, where $\rho$ is the plasma density, $\nu_{\rm t} \sim u \xi \sim \omega \xi^{2}$ the turbulent viscosity, and $e_{kh} = (1/2) (\partial u_{k}/\partial x_{h} + \partial u_{h} /\partial x_{k})$  the strain tensor with $u_{k}$ the velocity components and $x_{k}$ the Cartesian coordinates \citep[e.g.,][]{Chandrasekhar61}. The order of magnitude of the turbulent dissipation rate per unit volume is ${\cal D}  \sim 2 \rho u^{3}/\xi \sim 2 \rho \omega^{3} \xi^{2}$ because $u \sim \omega \xi$. It can be equated to the order of magnitude of the tidal power per unit volume that excites the oscillations, $\rho \omega \xi |\nabla \Psi_{lmn}|$, to estimate the amplitude $\xi$ in the stationary regime. In this way, we find
\begin{equation}
\xi \sim \frac{1}{2} \omega^{-2}  |\nabla \Psi_{lmn}|.
\label{ampl_turb}
\end{equation}
We applied Eq.~(\ref{ampl_turb}) to estimate the maximal amplitude of the oscillations in the turbulent regime in Sect.~\ref{applications}, finding typical values $\xi \ga 10^{6}$~m for several systems; this justifies the typical amplitude assumed to compute the amplification timescale in Eq.~(\ref{ampl_time}). 

\subsection{Tidal torque at resonance}
\label{tidal_torque_sec}

The expression of the power per unit volume at resonance given by Eq.~(\ref{res_power}) can be used to evaluate the total power dissipated inside a flux tube of volume $V_{\rm ft} =  (\pi^{2}/2) r_{0} d_{\rm ft}^{2}$ and the corresponding torque ${\cal T}$ acting on the stellar spin as 
\begin{equation}
{\cal T} = \frac{1}{\Omega_{\rm s}}\left[ -\rho_{0} \dot{\vec \xi} \cdot \nabla \Psi_{lmn} \right]_{\rm res} V_{\rm ft}. 
\label{tidal_torque}
\end{equation}
If the moment of inertia of the star is $I$, its angular velocity will change according to
\begin{equation}
\frac{d\Omega_{\rm s}}{dt} = \frac{\cal T}{I}. 
\label{dOmegadt}
\end{equation}
Therefore, the timescale for the variation in the stellar rotation is $\tau_{\Omega_{\rm s}} \equiv (d \log\Omega_{\rm s}/dt)^{-1}= I\Omega_{\rm s}/{\cal T}$. 

In Sect.~\ref{applications} we {make} use of Eq.~(\ref{tidal_torque}) to estimate $\tau_{\Omega_{\rm s}}$, finding values over a very wide range between $10^{5}$ and $10^{26}$ years for  the systems in Table~\ref{table1}.
Tidal timescales shorter than approximately $1$~Gyr are shorter than the timescale of evolution of the stellar rotation under the action of the magnetic braking by the stellar wind. Therefore, the tidal torque produced by the resonance can dominate the stellar spin evolution in such systems. On the other hand, longer tidal timescales correspond to systems where the resonant torque has virtually no effect on the evolution of the stellar rotation.
 
In the above estimate, we have assumed that the torque acts on the whole star. As a matter of fact, it acts on the slender flux tube because the tidal potential makes work on it to amplify or maintain its oscillations. Nevertheless, we assume that the above torque is rapidly redistributed over the whole star once the oscillations reach a sizeable amplitude. A description of this process and its effects on the dissipation of the kinetic energy of the flux tube oscillations are left for a future investigation because in the present model we do not include the coupling of the flux tube with the surrounding medium. As in other works considering the tidal excitation of oscillations in a star, we assume that the torque acts over the whole star because the exchanged angular momentum is redistributed in the stellar interior and is not limited to the layers where the waves are excited.

\subsection{Realization and maintenance of the resonance condition}
\label{maintaining_osc}

\subsubsection{Primary mechanism}

The rotation of the host star and the orbital period of the planet evolve on timescales on the order of $\sim 1$~Gyr because of the angular momentum loss in the stellar wind and the tides. The possible resonant frequencies are given by   combinations of integer values $(m,n)$ such that $\omega_{mn} = m\Omega_{\rm s} - n\Omega_{0}$ is sufficiently small and close to one of the zeros of the characteristic polynomial (i.e.,  $P(\tilde{\omega}_{mn}) \simeq 0$). Given the long-term evolution of the stellar rotation and of the planetary orbit, there are good chances that the resonance condition is matched at some time during the evolution of the system. 

The semiamplitude of the resonance, that is, the frequency difference $\Delta \tilde{\omega}_{\rm res}$ giving an amplitude of one half of  the maximum of $W$,  is given by the equation (cf. Eq.~\ref{w_eq}) 
\begin{equation}
\Delta \tilde{\omega}_{\rm res} = \frac{2 E \tilde{\zeta} \tilde{\omega}}{\left| P^{\prime}(\tilde{\omega}_{\rm res}) \right|}, 
\end{equation}
where $P^{\prime}(\tilde{\omega}_{\rm res})$ is the derivative of the characteristic polynomial at the resonant frequency. The semiamplitude can be expressed as a fraction of the stellar angular velocity as
\begin{equation}
\frac{\Delta \omega_{\rm res}}{\Omega_{\rm s}} \simeq \frac{2 E \tilde{\zeta}}{| 2 d_{2}\tilde{\omega}_{\rm res} + d_{1}| } \left( \frac{\omega_{\rm res}}{\Omega_{\rm s}}\right),
\label{resonance_width}
\end{equation} 
where we made use of the magnetostrophic approximation to neglect the $\tilde{\omega}^{4}$ term in the characteristic polynomial. Given that $\zeta$ is very small, the width of the resonance is extremely small, with typical values $\Delta \omega/\Omega_{\rm s} \sim 10^{-11} -10^{-9}$, as we shall see in Sect.~\ref{applications}. A flux tube with a constant magnetic field $B$ cannot be maintained inside such an extremely sharp resonance for a sufficiently long time as to excite a sizeable oscillation amplitude unless there is some mechanism that keeps  $B$ very close to the resonance value. 

\begin{figure}
%\hspace*{-7mm}
 \centering{
 \includegraphics[width=8cm,height=7cm,angle=0,trim=85 85 30 30,clip]{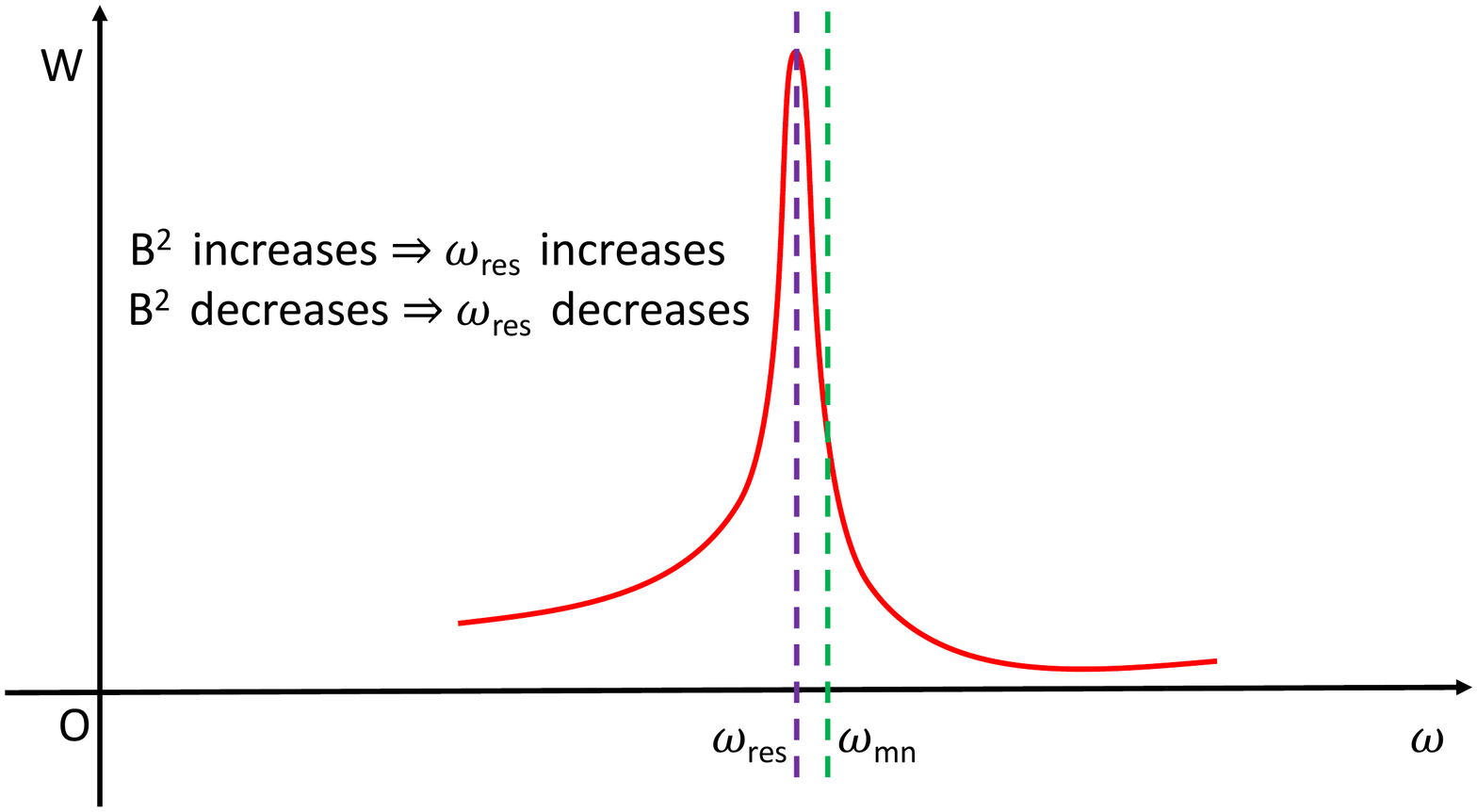}} % resonance_condition_beta.pdf}} % HD106252_1_rhkcorr_KR_final.pdf}} [trim=87 87 93 93, clip]
% \vspace*{-10mm}
   \caption{Resonance amplification factor, $W$, vs. frequency of oscillation, $\omega$ (solid red line). The vertical dashed violet line marks the resonance peak at $\omega=\omega_{\rm res}$, and the  dashed green line indicates the frequency of the tidal potential component that is responsible for exciting the oscillations,  $\omega_{mn}$. When the magnetic field, $B$, of the flux tube increases, the resonance peak  moves toward the right because $\omega_{\rm res}$ increases. On the contrary, when the magnetic field strength decreases, the peak moves toward the left. Therefore, the perturbation $\Delta B^{2}$ produced by the oscillation displacement, $\xi_{r}$, can lock the system into resonance, with $\omega_{mn}$ being slightly greater than $\omega_{\rm res}$, as illustrated (see text). } 
\label{resonance_plot_beta}%
\end{figure}

A possible mechanism may exploit the perturbation of the magnetic field strength produced by the resonant oscillations themselves.
The perturbation of the magnetic field $\vec B = B {\vec e}_{\phi}$ associated with the displacement $\vec \xi$, neglecting the magnetic field diffusion that operates on timescales much longer than the period of the oscillations, is given by the ideal induction equation of MHD, 
\begin{equation}
\delta {\vec B} = \nabla \times ({\vec \xi} \times {\vec B}) =  - \nabla \times [(\xi_{r} B) {\vec e}_{\theta}], 
\end{equation}
where ${\vec e}_{\theta}$ is the unit vector in the colatitude direction. Considering that $\partial {\vec \xi}/\partial \phi = j m \, {\vec \xi}$ and that $\xi$ does not depend on the radial coordinate in the slender flux tube approximation, we find
\begin{equation}
\delta {\vec  B} = \frac{B \xi_{r}}{r_{0}} \left( j m \, {\vec e}_{r} - {\vec e}_{\phi} \right),
\end{equation}
where ${\vec e}_{r}$ and ${\vec e}_{\phi}$ are unit vectors in the radial and azimuthal directions, respectively, and we made use of the hypothesis that the flux tube is in the equatorial plane. The variation in the square of the magnetic field that determines the parameter $\beta$ of our model is found to be 
\begin{equation}
\frac{\Delta B^{2}}{B^{2}} = - \frac{\Delta \beta}{\beta} = - \frac{m^{2}-1}{4} \frac{\xi_{r}^{2}}{r_{0}^{2}},
\label{resonance_maintain}
\end{equation}
where the factor $4$ in the denominator comes from the average over the azimuthal angle, $\phi_{0}$, and the time, which makes the first-order perturbation vanish. %\LEt{ or "the time that makes the first-order..." depending on your meaning.} 
As we shall see in Sect.~\ref{applications}, the frequency of resonance $\omega_{\rm res}$ increases when the magnetic field strength $B$ increases (or $\beta$ decreases). Therefore, if the frequency of the tidal potential responsible for the resonance $\omega_{mn}$ is slightly greater than the resonance frequency,  $\omega_{\rm res}$, and falls in the descending part of the resonance curve as shown in Fig.~\ref{resonance_plot_beta}, the resonance condition can be maintained despite any small fluctuation of the field. 

Specifically, if the field $B$ increases,  $\omega_{\rm res}$ will increase and the resonance peak will move toward the right making the amplification factor $W$ at the tidal frequency $\omega_{mn}$ larger. This will increase $\xi_{r}^{2}$ producing a negative perturbation of the square of the magnetic field for $m> 1$. Conversely, if there is a fluctuation that decreases $B$, the resonance peak will move to the left of $\omega_{mn}$ making $\xi_{r}^{2}$ smaller and producing a corresponding increase in $B$ for $m > 1$. In both the cases, the perturbation tends to restore the initial value of $B$ and of the difference $\omega_{mn}- \omega_{\rm res}$. 
%%%%%%%%%%%%%%%%%%%%%%%%%%%%%%%%%%%%%
%%%%%%%%%%%%%%%%%

The same self-regulating mechanism works also if the frequency of the tidal potential $\omega_{mn}$ varies. For example, if $\omega_{mn}$ decreases moving toward the peak of the resonance (see Fig.~\ref{resonance_plot_beta}), $\xi_{r}$ increases, the magnetic field perturbation becomes more negative, and the resonance peak moves toward lower frequencies, thus restoring the initial state. Similarly, if $\omega_{mn}$ increases, the amplification factor $W$ decreases making $\xi_{r}$ smaller and increasing the magnetic field $B$. In this way, the resonance peak moves toward higher frequencies again restoring the initial state. In this way, the proposed mechanism can maintain the fine tuning required to amplify the resonant oscillations for $m> 1$. 

The operation of this self-regulating mechanism requires a minimum amplitude of the radial oscillations, which can be estimated by considering that $\partial (\omega_{\rm res}/\Omega_{\rm s})/\partial B \sim 10^{-6}$~G$^{-1}$, as we shall find in Sect.~\ref{applications}. Considering that the full width at half maximum of the resonance peak is $ \la 10^{-9}$ in relative units, the magnetic field change to keep the system in resonance is $\Delta B /B \la 10^{-5}$ for a field intensity $B \sim 100$~G. Assuming a mean value $m=4$, the corresponding relative radial displacement is $\xi_{r}/r_{0} \la 2.3 \times 10^{-3}$, or  $\xi_{r} \la 10^{6}$~m in a Sun-like star ($r_{0} \sim 5 \times 10^{8}$~m), according to Eq.~(\ref{resonance_maintain}). Therefore, the possibility of exciting and maintaining the oscillations starting from an initial state of vanishing amplitude ($\xi_{r} \sim 0$) is questionable, but we assume that some descending convective plumes may penetrate from time to time into the overshoot layer and excite several normal modes of oscillations in the flux tube. If one such mode has an amplitude of $\sim 10^{6}$~m and its frequency resonates with the tidal potential, its amplitude will be maintained against the slow viscous decay and it can be locked into resonance by the mechanism described above. In other words, the proposed mechanism can work if the timescale $\Delta t$ given by Eq.~(\ref{ampl_time}) to excite the oscillations is comparable to or shorter than the timescale of viscous decay that we estimated to be $\zeta^{-1} \sim 10^{7}$ yr in Sect.~\ref{internal_m_field}. 

The evolution of the stellar spin and of the orbit of the planet remarkably changes $\omega_{mn}$ over a typical timescale of $10^{8}-10^{9}$ yr. The crossing of the extremely narrow resonance peak as a consequence of that evolution requires a time interval on the order of only one year during which the above considered penetrating plumes must be capable of triggering the resonant oscillations. The chances of  triggering can be greatly increased because the angular velocity of rotation in the overshoot layer may oscillate on different timescales as found, for example, in the case of the Sun. Variations with a period of $\sim 1.3$~yr have a relative semiamplitude of $\Delta \Omega_{\rm s}/\Omega_{\rm s} \sim 3 \times 10^{-3}$ and may not be a persistent feature \citep{Howe09}. Their frequency is significantly greater than $\omega_{mn}$, and thus their effect is averaged out for our purpose. Nevertheless, we may speculate that other modulations of the internal rotation with comparable amplitudes but periods of several decades or centuries may take place. In this case, the crossing of the resonance peak will not happen just once for all, but it will occur many times over a time interval of $10^{5}-10^{6}$ yr giving many opportunities to the infrequently penetrating convective plumes of triggering the oscillations in our flux tube in the overshoot layer. 

\subsubsection{Secondary mechanisms helping to maintain the system in resonance}
\label{secondary_reg_mechanisms}

In addition to the primary mechanism proposed above, we envisage two possible consequential mechanisms that may act to keep the system close to resonance. First, we consider the case when the additional tidal torque produced by the resonant oscillations (see Sect.~\ref{tidal_torque_sec}) is stronger than the  torque produced by the magnetized stellar wind  and the rotation period of the star is longer than the orbital period, so that the tidal torque acts to spin up the star, while the wind torque acts to slow it down. 

\begin{figure}
%\hspace*{-7mm}
 \centering{
 \includegraphics[width=8cm,height=7cm,angle=0,trim=85 85 30 30,clip]{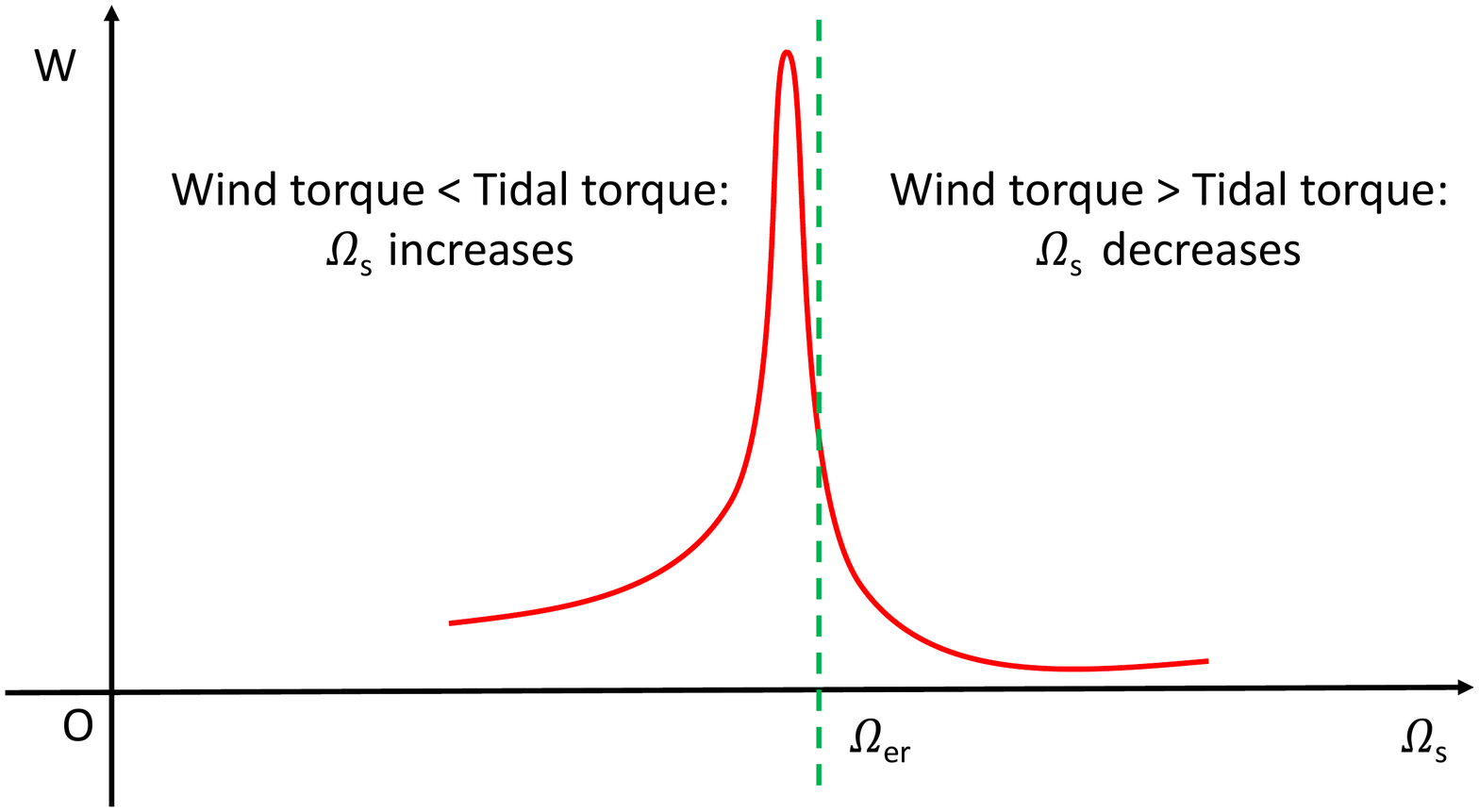}} % resonance_condition_new.pdf}} % HD106252_1_rhkcorr_KR_final.pdf}} [trim=87 87 93 93, clip]
% \vspace*{-10mm}
   \caption{Resonance amplification factor, $W$, vs. stellar angular velocity, $\Omega_{\rm s}$ (solid red line). The vertical dashed green line indicates the value of the stellar angular velocity, $\Omega_{\rm er}$, which is stably maintained by the opposite effects of the stellar wind torque, acting to brake stellar rotation, and the value of the resonant tidal torque introduced in Sect.~\ref{tidal_torque_sec}, acting to spin up the star when the orbital period of the planet is shorter than the stellar rotation period (see text). } 
\label{resonance_plot}%
\end{figure}

When the spin angular velocity, $\Omega_{\rm s}$, and the mean orbital motion, $\Omega_{0}$, come sufficiently close to the resonance, the amplitude of the flux tube oscillations starts to increase. In this regime, an increasingly stronger and stronger torque acts on the star, tending to increase its angular velocity faster than the braking by the wind. We speculate that the two opposite torques, that is, the wind torque and the tidal torque coming from the oscillations, act together to keep the values of $\Omega_{\rm s}$ immediately beyond the resonance peak on the descending branch of the resonance curve (cf. Fig.~\ref{resonance_plot}). If, for example, $\Omega_{\rm s}$ increases beyond the equilibrium value $\Omega_{\rm er}$ and moves away from the resonant peak, the tidal torque will decrease and the wind torque will take the lead, slowing down the stellar spin until the resonance condition is recovered. Conversely, if $\Omega_{\rm s}$  decreases below $\Omega_{\rm er}$, the system will move toward the resonance peak and the tidal torque will increase and accelerate the stellar spin, recovering the initial state.  In this way, the system will be kept close to the resonance. Only on an evolutionary timescale will a loss of the resonant magnetic flux tubes stored in the overshoot layer, a tidal decay of the planetary orbit, or changes in the depth of the stellar convection zone drive the system out of resonance, allowing it to resume its unperturbed spin-orbit evolution. %\LEt{ Verify that your intended meaning has not been changed.}
%%%%%%%%%%%%%%%%%%%%%%%%%%%%%%%%%%%%%
%%%%%%%%%%%%%%%%%

A second mechanism can be envisaged by considering that  the resonant oscillations may induce the emergence of stellar magnetic fields to the photosphere. The active regions observed in the photosphere of the Sun and solar-like stars have a latitudinal distribution and a tilting that can be explained by considering flux tubes with a strong field  ($\ga 10^{5}$~G) amplified and stored in the overshoot layer \citep{Caligarietal95,Granzeretal00}. We can speculate that the velocity field of the oscillations of the toroidal flux tube considered in the present model, which has a weaker magnetic field ($\la 10^{3} - 10^{4}$~G), {propagates in the low-viscosity medium that surrounds it in the overshoot layer and perturbs the strong-field flux tubes by inducing small deformations in them with a displacement pattern that has the same azimuthal wavenumber, $m$, as the resonant oscillations.  %\LEt{ Verify that your intended meaning has not been changed.}
The first modes to become unstable in those strong-field flux tubes are those with $m=1$ or $m=2$ \citep{Granzeretal00,Fan21}, implying that each strong flux tube may reproduce and amplify one or two of the crests of the oscillations of the underlying resonant flux tube.} When these crests of the perturbed strong flux tubes enter the convection zone, they become strongly unstable and magnetic buoyancy rapidly accelerates them toward the surface of the star, provided that the horizontal wavelength of the perturbations is greater than twice the local pressure scale height, which is well verified for $m \la 20$ \citep[][Ch. 8.2]{Priest84}. 

In this way, a pattern of surface spots may be formed that rotates with the angular velocity of the oscillating flux tube and whose distribution in longitude replicates the azimuthal wavenumber of its oscillations. For a distant observer, such starspots rotate in close commensurability with the orbital motion of the planet, which can account for the observations presented in Sect.~\ref{observations}. 
Moreover, this spot pattern affects the large-scale configuration of the surface magnetic field of the star making it dominated by a mode with a rather large $m$ corresponding to the azimuthal  wavenumber of the flux tube oscillations. This may have consequences for the wind braking because a potential field configuration with a rather large $m$ must have a degree $l \geq |m|$, which implies the predominance of high-order multipoles in the stellar coronal field. In this high-$l$ regime, the angular momentum loss rate by the wind is strongly reduced \citep[e.g.,][]{Revilleetal15}, thus slowing down the evolution of stellar rotation, an effect that can contribute to the fine-tuning required by the resonance for rather long time intervals. 
 
This mechanism does not imply a self-regulating process as in the case of the other mechanisms, but it increases the time required to cross the resonance because it slows down the evolution of the stellar rotation. It may act also in the case of stars rotating faster than the planetary orbit when the tidal interaction is small because of the relatively large $a/R$ or the small mass of the planet. Isolated stars of spectral type A or F experience a weaker wind braking because of their thin convection zones, supporting a dynamo that is weaker than those of G or K stars \citep[e.g.,][]{WolffSimon97}. Therefore, in the case of A or F stars,  a resonance accompanied by the emergence of $m$ crests of the azimuthal field may make their spin braking even weaker, thus allowing the system to cross the resonance very slowly. This increases the probability of observing them in a resonant regime. Among the systems listed in Table~\ref{table1} those with $P_{\rm rot} < P_{\rm orb}$ have $a/R \ga 18$ or $T_{\rm eff} \ga 6100$~K with the only exception of AU Mic. Therefore, they may be in a state of highly reduced wind braking, that is, in a phase of stalled rotational evolution. 
 
The case of AU Mic can be peculiar because it is still in its pre-main-sequence phase and is contracting toward its final radius on the main sequence. Therefore, we may speculate that the resonance is happening during a phase when the braking by its intense wind is almost exactly counterbalanced by the spin up produced by the reduction of its moment of inertia. In this regime, the star behaves similarly to the early-type hosts considered above and it could be possible to explain the slowing down of its rotational evolution at resonance with the same mechanism. 

\subsection{Oscillations in the latitudinal direction}
Before leaving this section dedicated to the theory, a consideration of the forced oscillations in latitude described by Eq.~(\ref{eq_motion3}) is in order. They can resonate with a periodically oscillating tidal force in the latitudinal direction at a frequency
\begin{equation}
\tilde{\omega}_{\theta \, \rm res} \simeq f \sqrt{2 m^{2} -1}, 
\end{equation}
where we have neglected the small difference $x_{\rm e} -x$ in the numerator and $x_{\rm e}$ in the denominator in the term proportional to $\xi_{\theta}$ on the right hand side of Eq.~(\ref{eq_motion3}) that produce a variation of less than 5\%. The resonant frequency, in physical units, is $\tau^{-1} \tilde{\omega}_{\theta \, \rm res} \sim 6.1 \times 10^{-8}$~s$^{-1}$, that is, about 6\% of the spin frequency of a star with $P_{\rm rot} \sim 8$~days. In other words, no tight spin-orbit commensurability can be expected in such a case. Therefore, we did not consider latitudinal resonant oscillations as a possible explanation for the observed spin-orbit commensurability and concentrated only on the oscillations in the equatorial plane. 

\section{Applications}
\label{applications}
%The resonance frequencies are given by the solution of the characteristic polynomal equation (\ref{char_eq}). They have been studied by  \citet{FerrizMasSchussler94} to whom we refer the reader  for a general discussion, focusing here on the solution in the low-frequency magnetostrophic regime (see Eq.~\ref{magnetostrophic_sol}). 

The solution~(\ref{magnetostrophic_sol}) of the characteristic equation giving the resonant frequency depends on the three real coefficients $d_{2}$, $d_{1}$, and $d_{0}$. In turn, they are functions of: the parameters $\sigma$, $f$, $\delta$, $q$, and $\Omega_{\rm s}$, which depend on the stellar interior structure and rotation; the radius of the flux tube, $r_{0}$, which we assume equal to that of the overshoot layer where it is located; and $m$, $\beta$, $\Omega_{\rm s}$, and $\Omega_{\rm e0}/\Omega_{\rm s}$, which depend on the azimuthal wavenumber, $m$, of the considered oscillation mode and the properties of the magnetic  flux tube.  The ratio $\Omega_{\rm e0}/\Omega_{\rm s}$ is constrained by the requirement of a stable flow inside the flux tube against the Kelvin-Helmholtz instability that requires $r_{0} | \Omega_{\rm s} - \Omega_{\rm e0} | \leq v_{\rm A} $ \citep[see][ their Eq.~4.11]{FerrizMasSchussler93}. We consider the limit case when $\Omega_{\rm e0} = \Omega_{\rm s} - v_{\rm A}/r_{0}$  to be sure that our assumption does not affect the conclusions on the stability of the flux tube in the absence of external forcing \citep[cf. ][]{FerrizMasSchussler93}. The stratification in the overshoot layer is poorly known; therefore, we explored different values of the subadiabatic gradient, $\delta$. Similarly, the strength of the magnetic field $B$ is a priori unknown, leading us to explore a wide range, at least between 10 and 1000~G. 

In our analysis, we fix the stellar internal stratification and the rotation giving $\sigma$, $\Omega_{\rm s}$, and $q=0.06$ (the solar value), and explore different values of the magnetic field strength inside the flux tube $B$ and  of the subadiabatic gradient $\delta$. 

We make use of MESA stellar evolution models \citep[][]{Fieldsetal18,Paxtonetal19}, made accessible through the MESA-web interface\footnote{See http://www.astro.wisc.edu/~townsend/static.php?ref=mesa-web}, to compute the internal pressure, density, pressure scale height $H$, $\sigma$, and the radius $r_{0}$ assumed equal to the base radius of the external convection zone according to the standard Schwarzschild criterion with a ratio of the mixing length to the local pressure scale height equal to 2.0 with the Cox prescription for the mixing length theory and no overshooting. For the computation of our stellar models, we assume solar chemical abundances, no rotation, and the basic set of nuclear reactions and opacities provided in the code implementation through the website. For each of the stars listed in Table~\ref{table1}, we chose the internal structure model closer in mass and radius to those of the target, independently of its age, provided that the model corresponds to the main-sequence evolution phase. Only for AU~Mic we chose a pre-main-sequence model, because of its young age ($\approx 23$~Myr). 

\begin{figure}
%\hspace*{-7mm}
 \centering{\hspace*{-2cm}
 \includegraphics[width=9cm,height=10.5cm,angle=90]{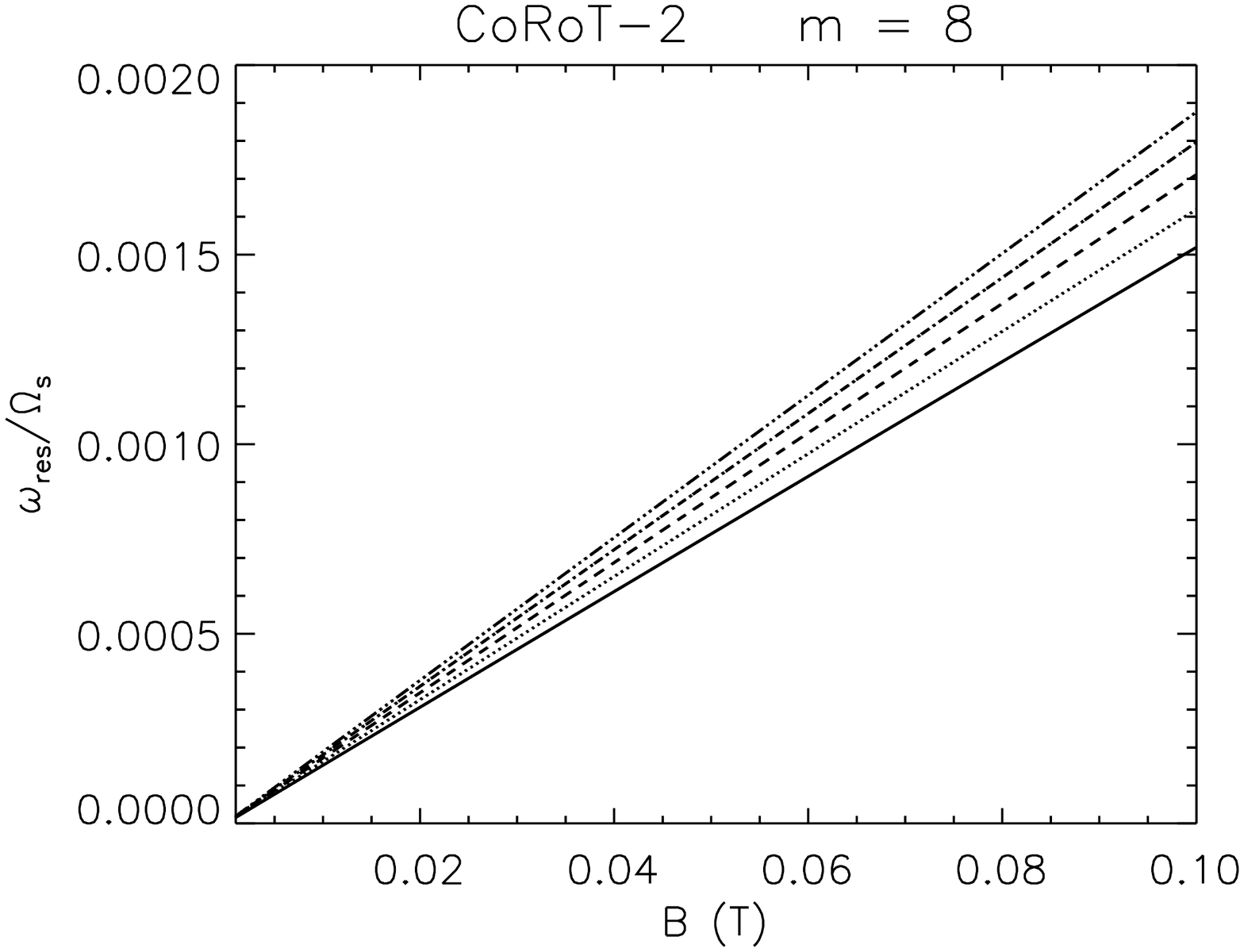} % CoRoT-2_ref_corr.pdf}% HD106252_1_rhkcorr_KR_final.pdf}} [trim=87 87 93 93, clip]
% \vspace*{-10mm}
   \caption{Resonant frequency, $\omega_{\rm res}$, in units of the stellar spin frequency, $\Omega_{\rm s}$, for CoRoT-2 vs. the magnetic field strength, $B$, of the toroidal flux tube in its overshoot layer as given by Eq.~(\ref{magnetostrophic_sol}) for different values of the subadiabatic gradient: $\delta = -6 \times 10^{-6}$ (three-dot-dashed line), $\delta=-5 \times 10^{-6}$ (dot-dashed line), $\delta = -4 \times 10^{-6}$ (dashed line),  $\delta = -3 \times 10^{-6}$ (dotted line), and $\delta = -2 \times 10^{-6}$ (solid line). }
   \label{corot2_plot}}
%\label{corot2_plot}%
\end{figure}

\begin{figure}
%\hspace*{-7mm}
 \centering{\hspace*{-2cm}
 \includegraphics[width=9cm,height=10.5cm,angle=90]{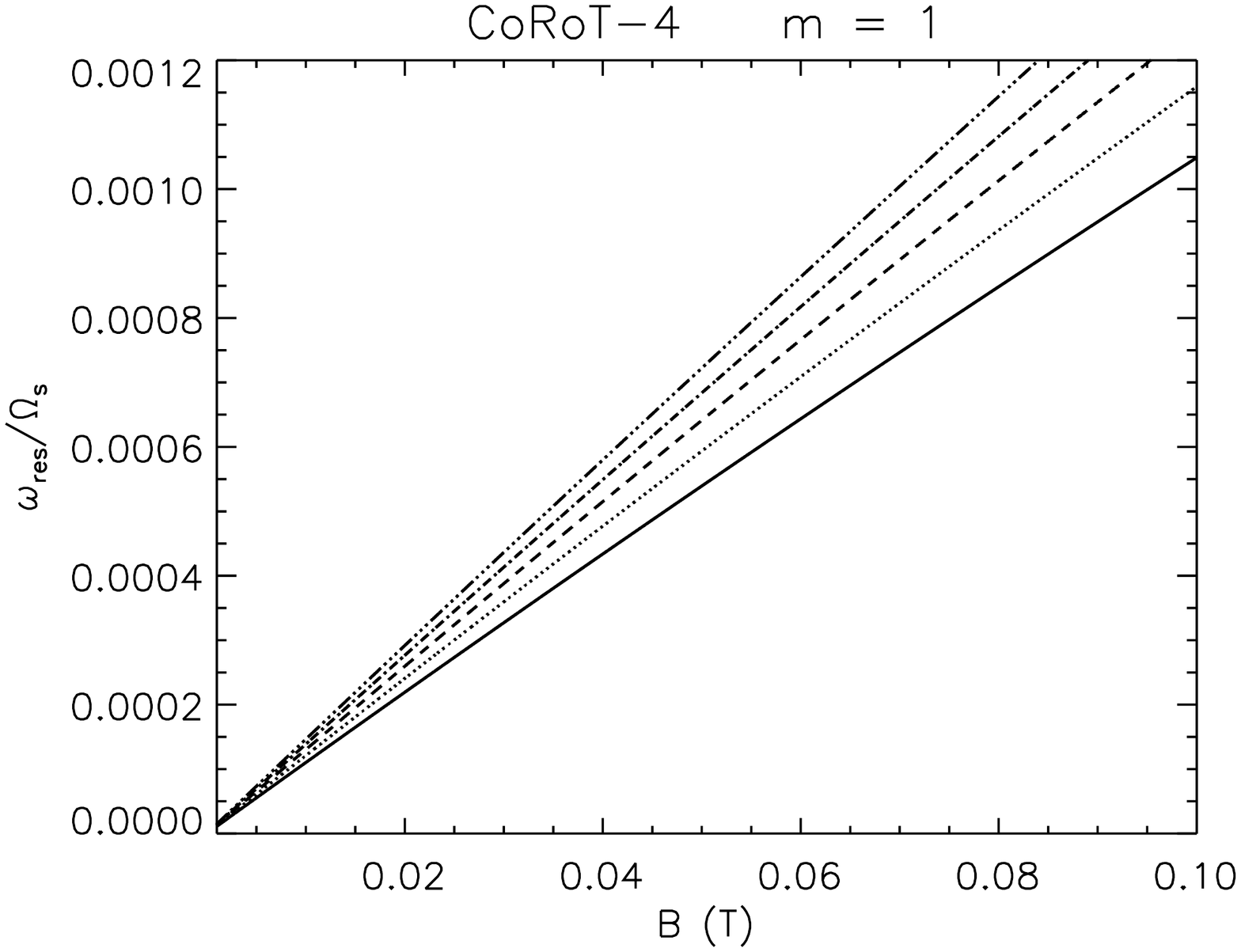} % CoRoT-4.pdf}% HD106252_1_rhkcorr_KR_final.pdf}} [trim=87 87 93 93, clip]
% \vspace*{-10mm}
   \caption{Same as Fig.~\ref{corot2_plot}, but for CoRoT-4. }}
\label{corot4_plot}%
\end{figure}

\begin{figure}
%\hspace*{-7mm}
 \centering{\hspace*{-2cm}
 \includegraphics[width=9cm,height=10.5cm,angle=90]{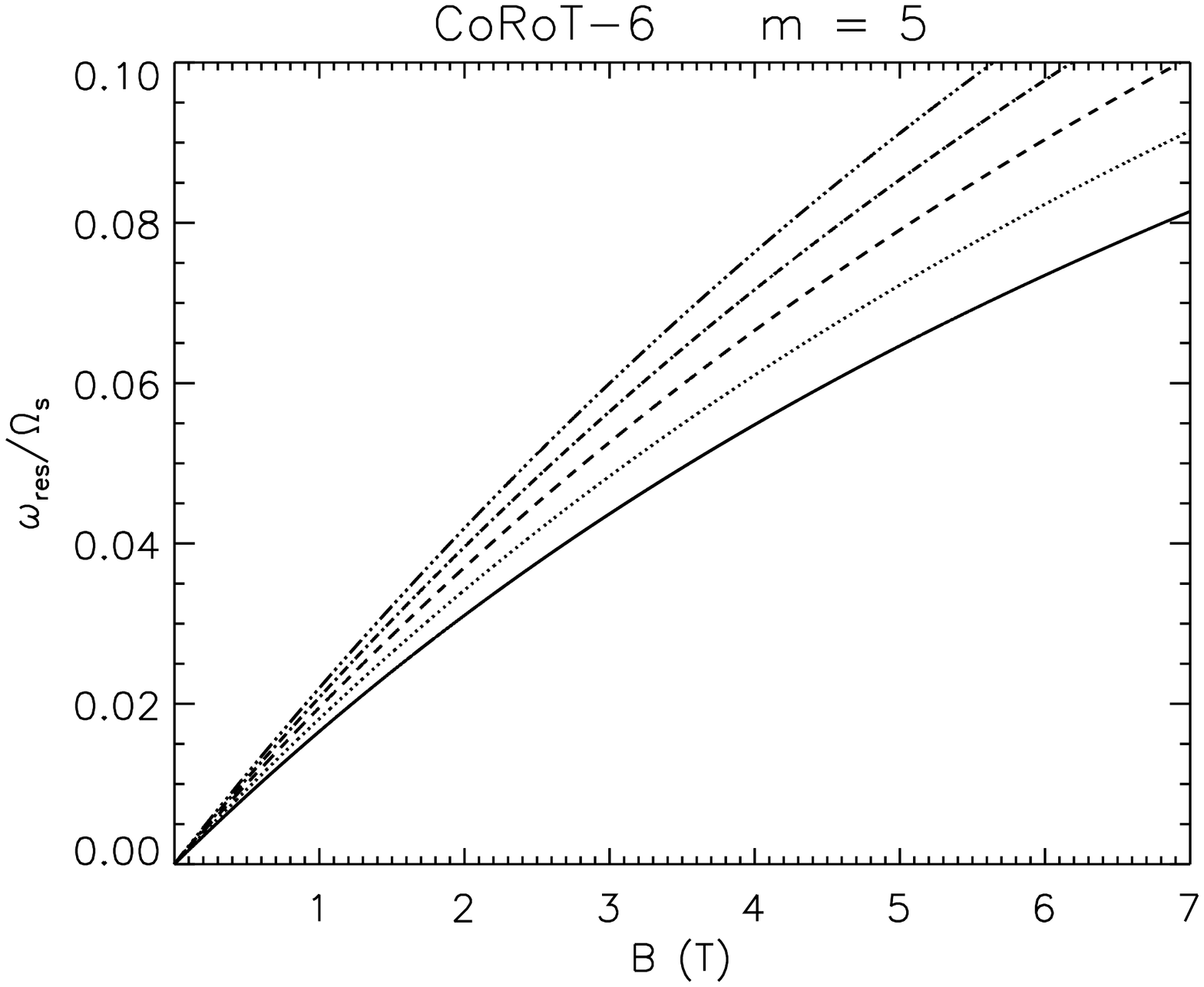} % CoRoT-6_strong_field.pdf}% HD106252_1_rhkcorr_KR_final.pdf}} [trim=87 87 93 93, clip]
% \vspace*{-10mm}
   \caption{Same as Fig.~\ref{corot2_plot}, but for CoRoT-6. Note the extended range of the magnetic field in comparison to the case of CoRoT-2. } \label{corot6_plot}}
\end{figure}

\begin{figure}
%\hspace*{-7mm}
 \centering{\hspace*{-2cm}
 \includegraphics[width=9cm,height=10.5cm,angle=90]{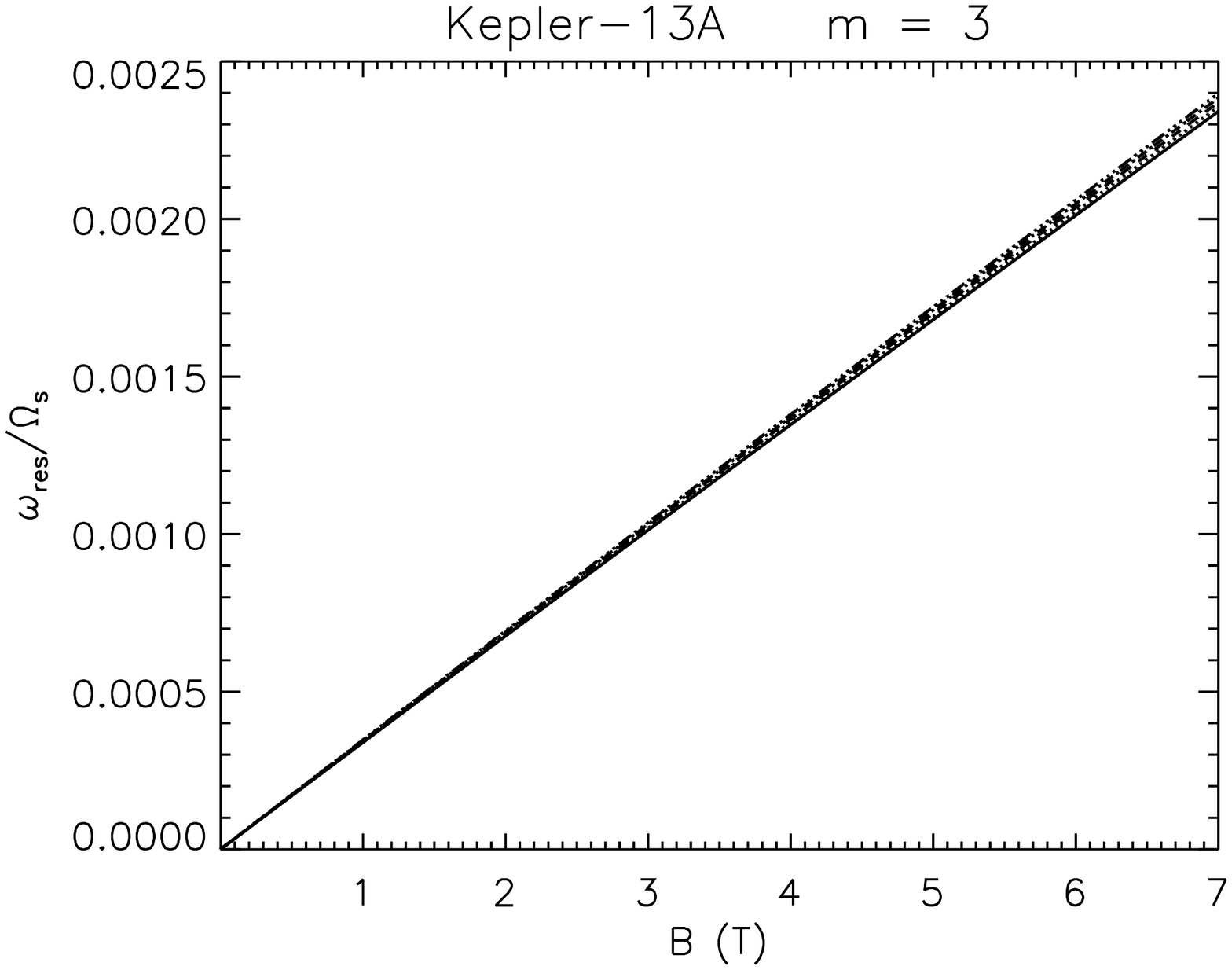} % Kepler-13A_strong_field.pdf}% HD106252_1_rhkcorr_KR_final.pdf}} [trim=87 87 93 93, clip]
% \vspace*{-10mm}
   \caption{Same as Fig.~\ref{corot2_plot}, but for Kepler-13A. Note the extended range of the magnetic field in comparison to the case of CoRoT-2.}}
\label{kepler13A_plot}%
\end{figure}

\begin{figure}
%\hspace*{-7mm}
 \centering{\hspace*{-2cm}
 \includegraphics[width=9cm,height=10.5cm,angle=90]{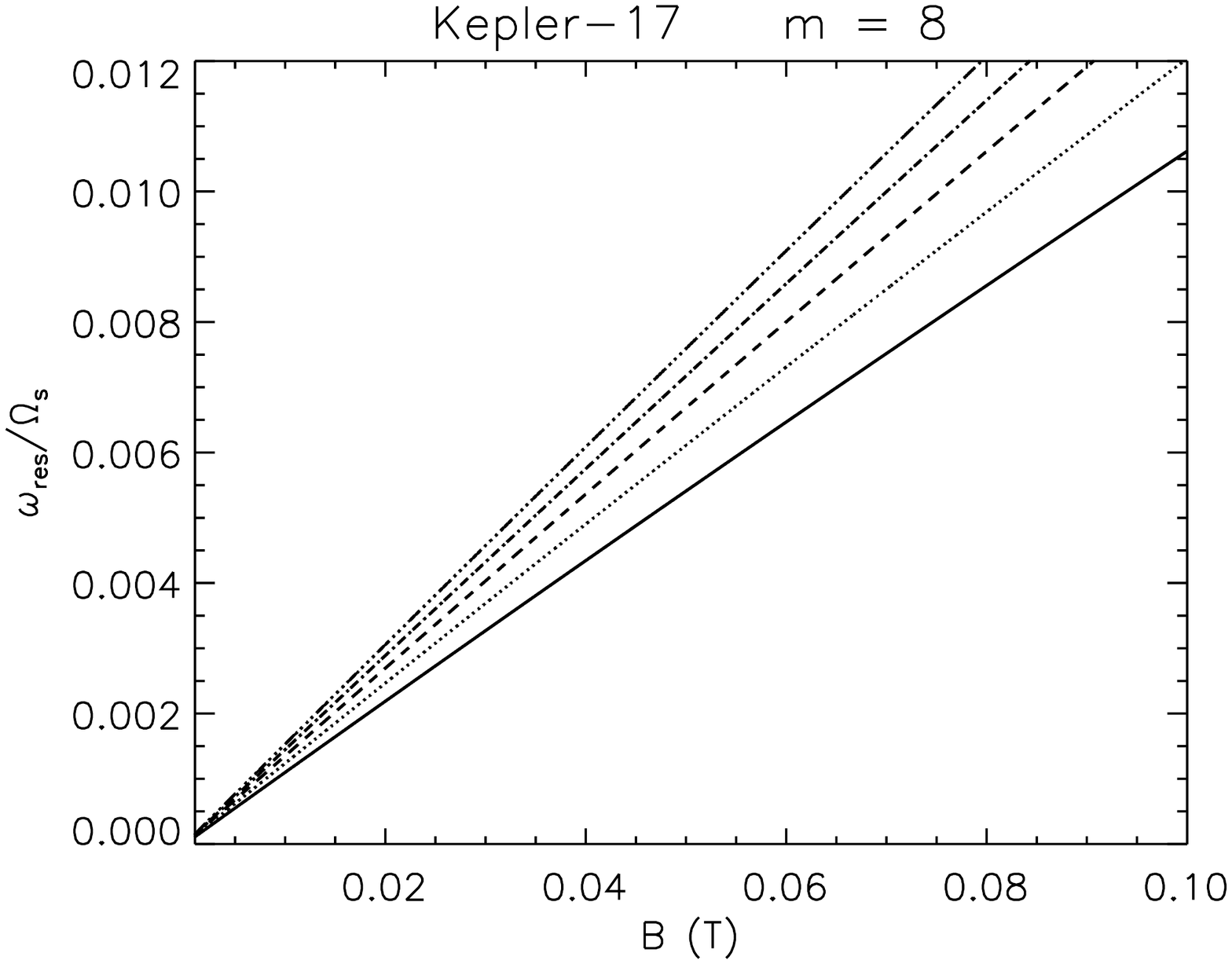} % Kepler-17.pdf}% HD106252_1_rhkcorr_KR_final.pdf}} [trim=87 87 93 93, clip]
% \vspace*{-10mm}
   \caption{Same as Fig.~\ref{corot2_plot}, but for Kepler-17. }}
\label{kepler17_plot}%
\end{figure}

\begin{figure}
%\hspace*{-7mm}
 \centering{\hspace*{-2cm}
 \includegraphics[width=9cm,height=10.5cm,angle=90]{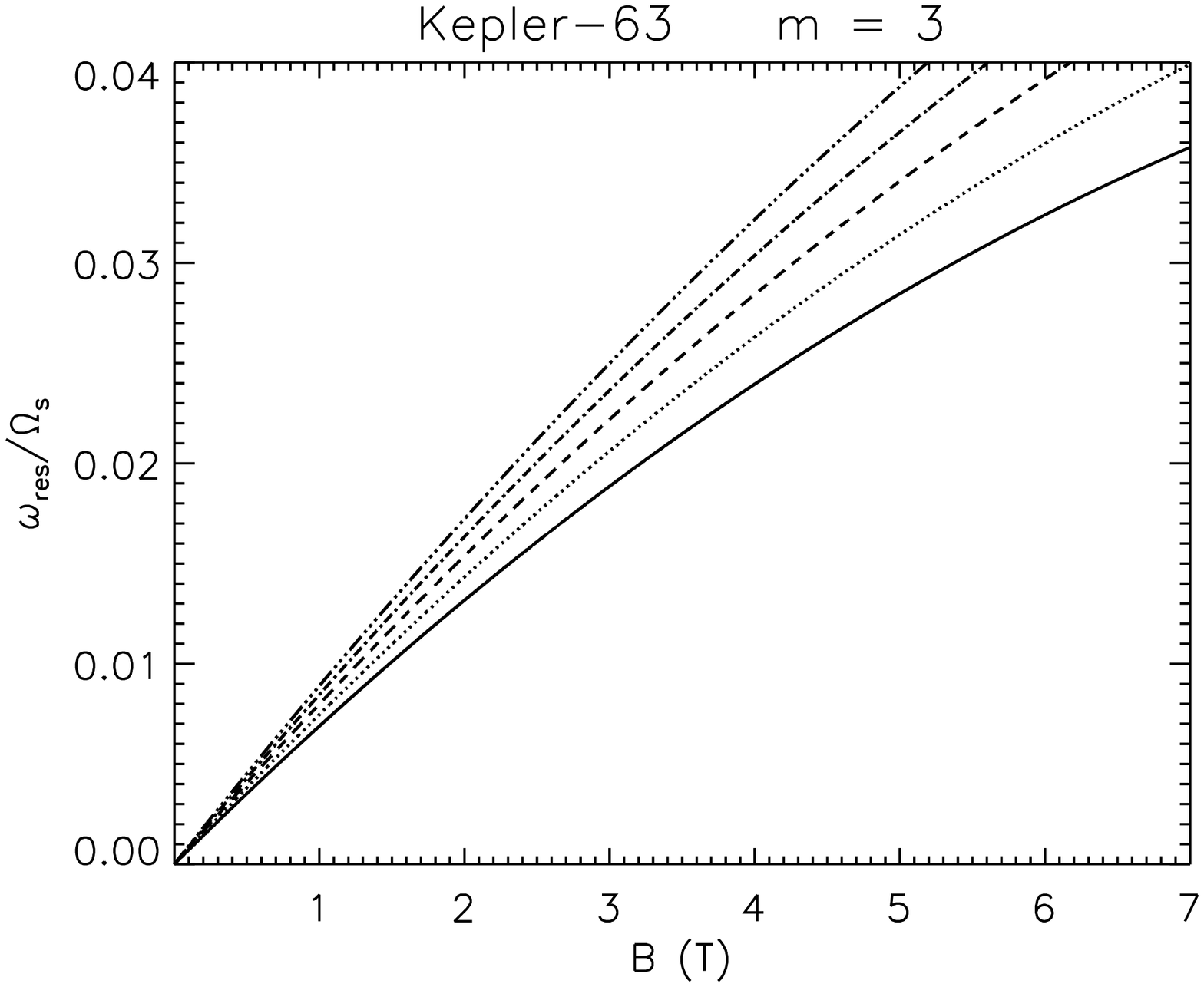} % Kepler-63_strong_field.pdf}% HD106252_1_rhkcorr_KR_final.pdf}} [trim=87 87 93 93, clip]
% \vspace*{-10mm}
   \caption{Same as Fig.~\ref{corot2_plot}, but for Kepler-63. Note the extended range of the magnetic field in comparison to the case of CoRoT-2. } \label{kepler63_plot}}
%\label{kepler63_plot}%
\end{figure}

\begin{figure}
%\hspace*{-7mm}
 \centering{\hspace*{-2cm}
 \includegraphics[width=9cm,height=10.5cm,angle=90]{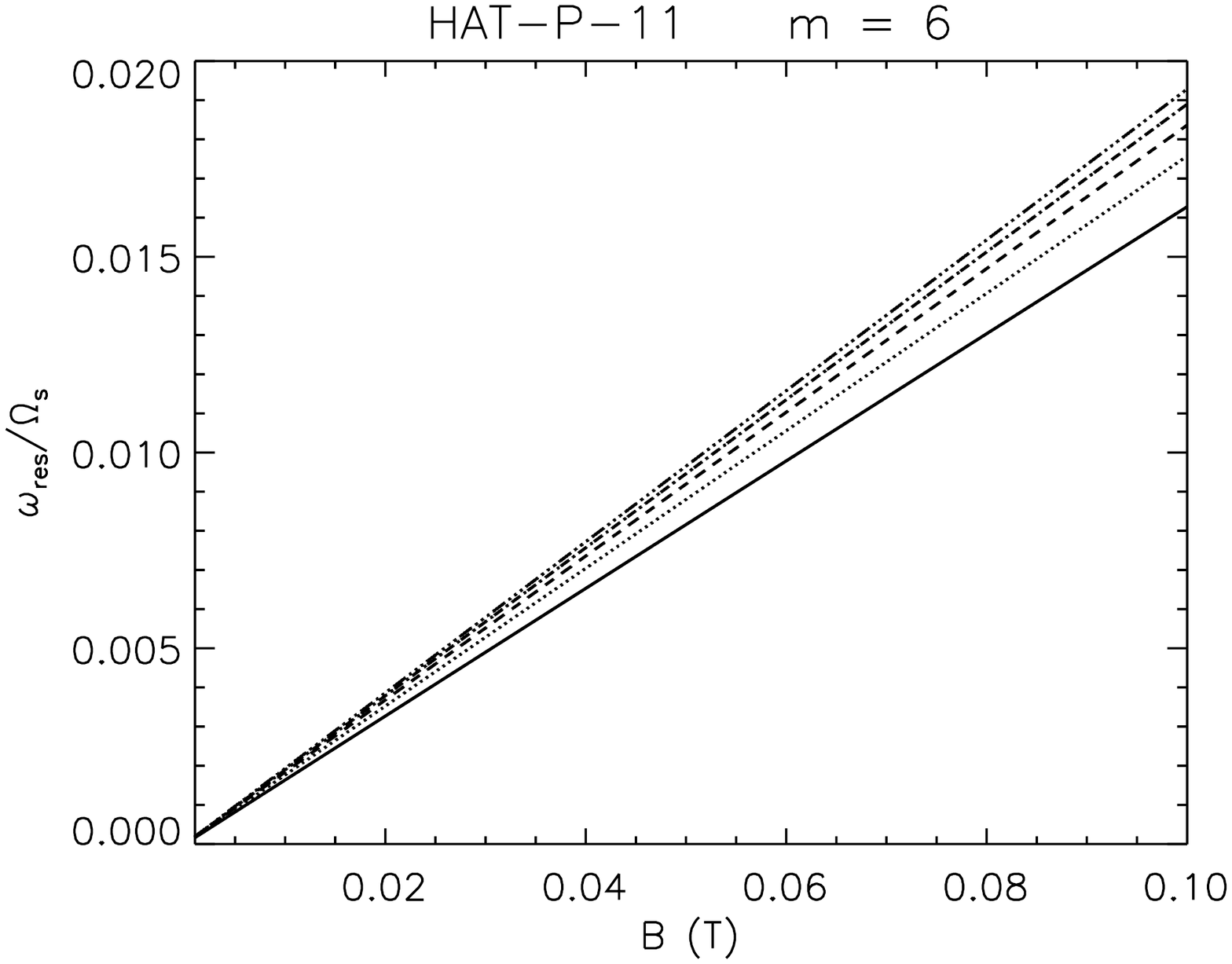} % HAT-P-11.pdf}% HD106252_1_rhkcorr_KR_final.pdf}} [trim=87 87 93 93, clip]
% \vspace*{-10mm}
   \caption{Same as Fig.~\ref{corot2_plot}, but for HAT-P-11. }}
\label{hat-p-11_plot}%
\end{figure}

\begin{figure}
%\hspace*{-7mm}
 \centering{\hspace*{-2cm}
 \includegraphics[width=9cm,height=10.5cm,angle=90]{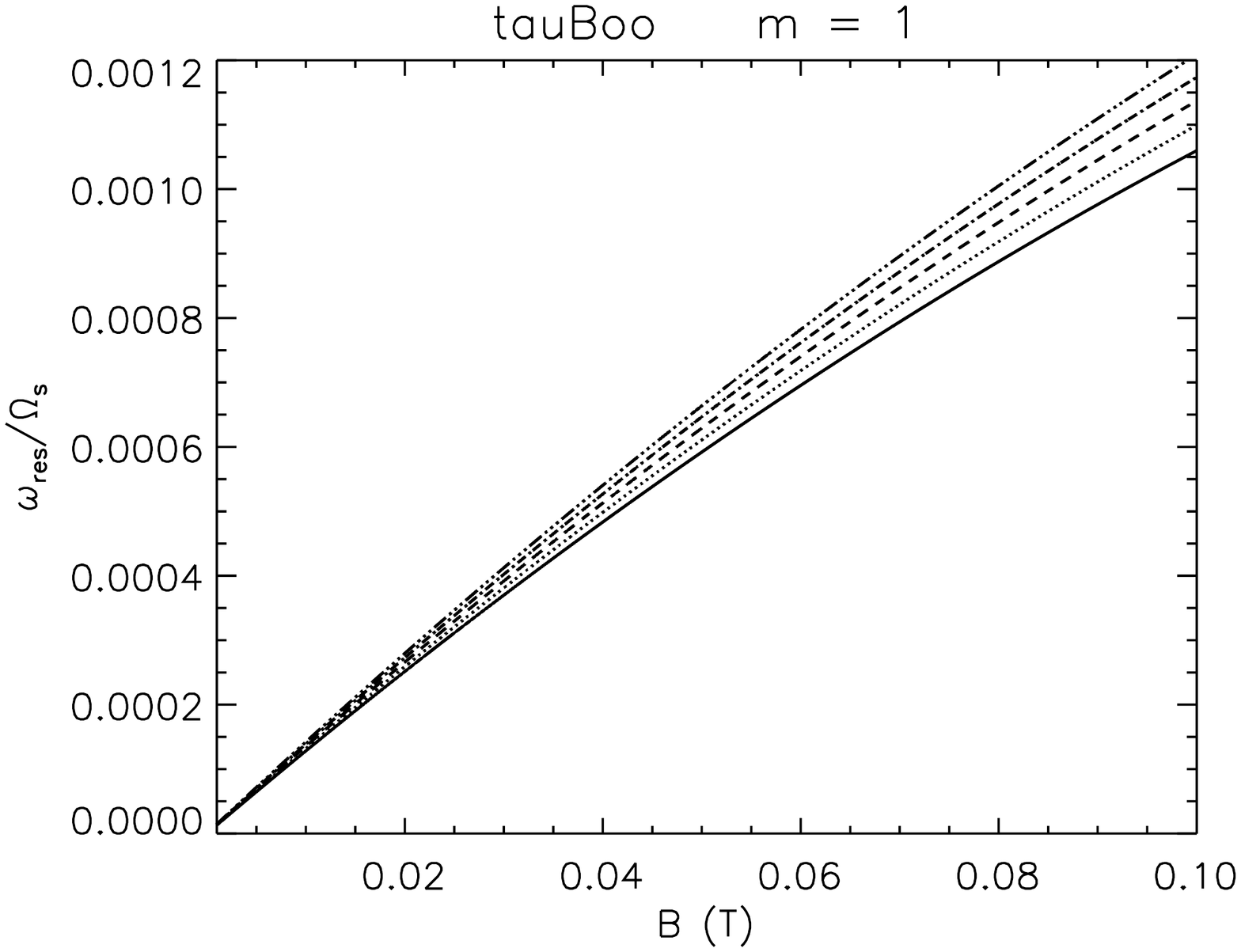} % tauBoo.pdf}% HD106252_1_rhkcorr_KR_final.pdf}} [trim=87 87 93 93, clip]
% \vspace*{-10mm}
   \caption{Same as Fig.~\ref{corot2_plot}, but for $\tau$~Boo. }}
\label{tauBoo_plot}%
\end{figure}

\begin{figure}
%\hspace*{-7mm}
 \centering{\hspace*{-2cm}
 \includegraphics[width=9cm,height=10.5cm,angle=90]{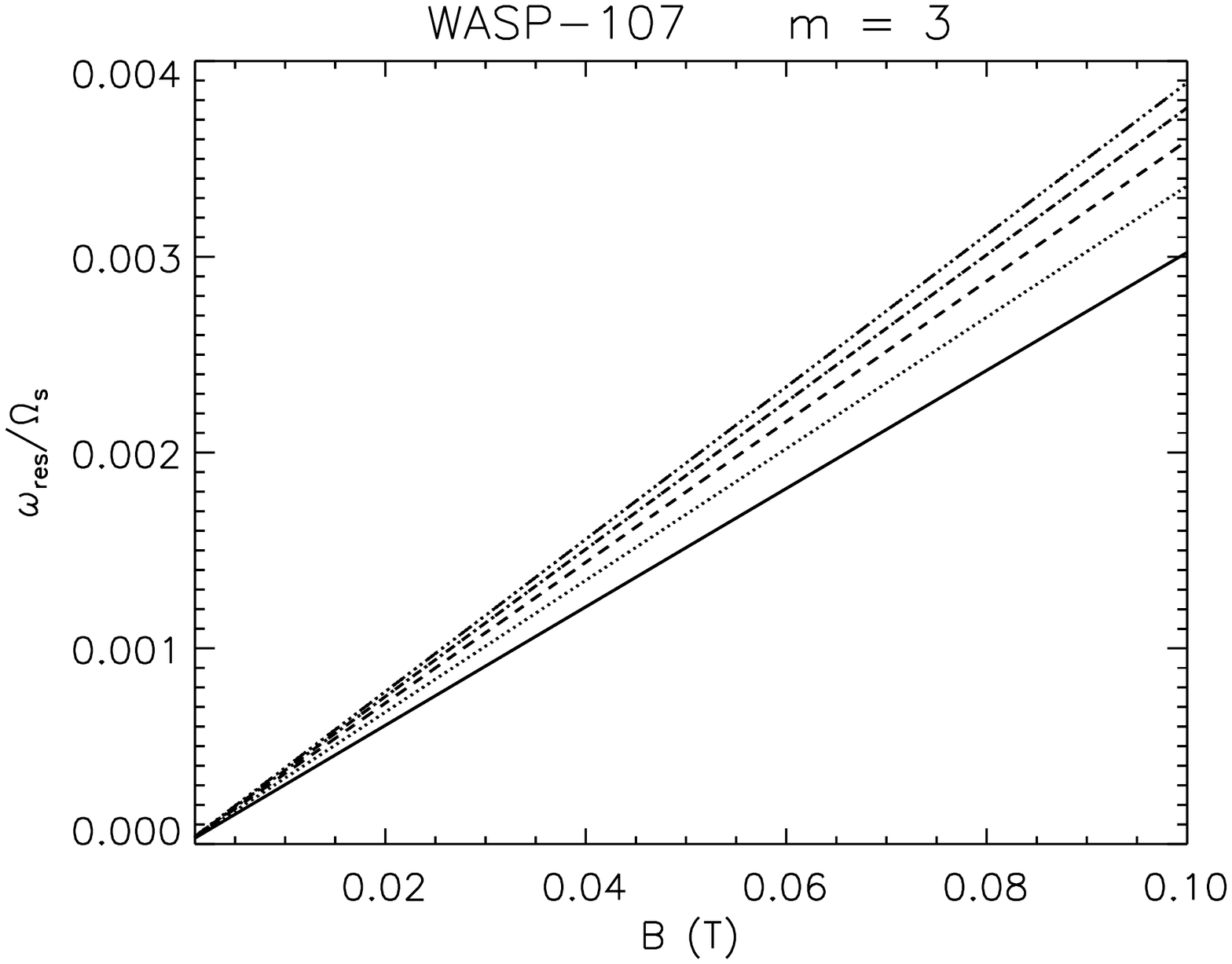} % WASP-107.pdf}% HD106252_1_rhkcorr_KR_final.pdf}} [trim=87 87 93 93, clip]
% \vspace*{-10mm}
   \caption{Same as Fig.~\ref{corot2_plot}, but for WASP-107.  }}
\label{wasp107_plot}%
\end{figure}

\begin{figure}
%\hspace*{-7mm}
 \centering{\hspace*{-2cm}
 \includegraphics[width=9cm,height=10.5cm,angle=90]{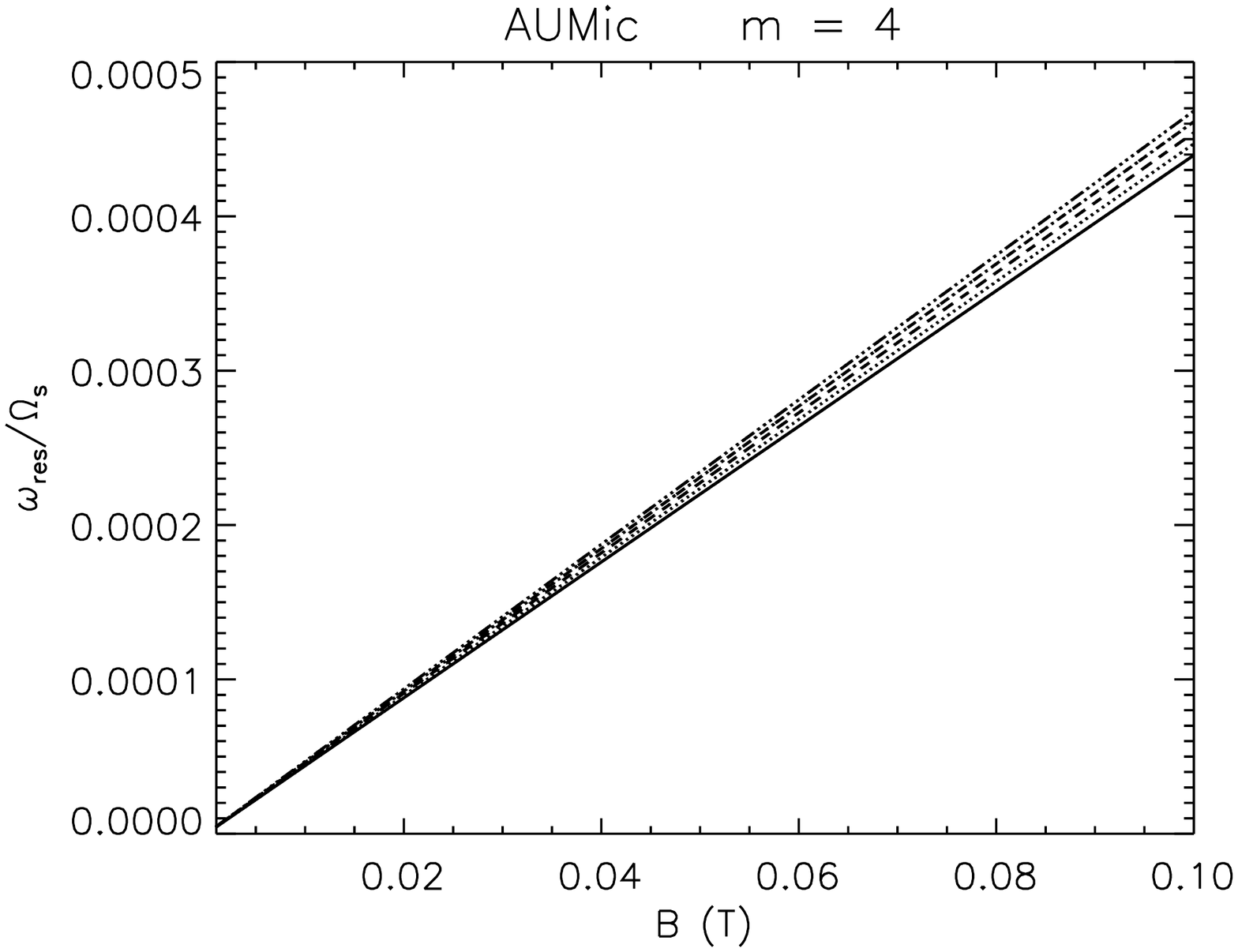} % AUMic.pdf}% HD106252_1_rhkcorr_KR_final.pdf}} [trim=87 87 93 93, clip]
% \vspace*{-10mm}
   \caption{Same as Fig.~\ref{corot2_plot}, but for AU~Mic. } 
\label{fig_aumic_plot_a}}%
\end{figure}

The resonant frequency $\omega_{\rm res}$ versus  the magnetic field strength $B$ in the flux tube is plotted for each of our targets in Figs.~\ref{corot2_plot} to~\ref{fig_aumic_plot_a} for a subadiabatic gradient $\delta$ ranging from $-6\times 10^{-6}$ to $-2\times 10^{-6}$. The values of $m$ and $n$ listed in Table~\ref{table1} are adopted together with the minimum possible $l \geq m$ giving $l-n$ even. The obliquity of all the systems is assumed to be $20^{\circ}$, except for Kepler-13A ($i=25^{\circ}$), Kepler-63 ($i = -120^{\circ}$), HAT-P-11 ($i = 70^{\circ}$), and WASP-107 ($i=40^{\circ}$). Such values are derived from the measurements of the projected spin-orbit misalignment $\lambda$ as obtained from the Rossiter-McLaughlin effect or starspot occultations (see Sect.~\ref{observations} and Table~\ref{table1}). 

The parameters determining the tidal potential component $\Psi_{lmn}$ that excites the resonant oscillations in our model are listed in Table~\ref{table_tidal_numbers}, where we report for each system the integers $l,m,n$ together with the absolute value of the coefficient $A_{lmn} (0,i)$ corresponding to the above obliquities. For simplicity, we consider in all the cases circular planetary orbits neglecting the effects of the eccentricity that does not greatly affect the order of magnitude of the tidal potential components. 
%%%%%%%%%%%%%%%%%%%%%%%%%
\begin{table}
\caption{Tidal potential parameters for our systems.}
\begin{center}
\begin{tabular}{ccccc}       
\hline                                                                                   
                      System  & $l$ & $m$ & $n$ & $\left| A_{lmn}(0,i) \right| $ \\
                      \hline 
                        CoRoT-2& 9 & 8 &   3 &   5.0489e-03\\
                        CoRoT-4 &   3 &   1 &   1 &      4.0068e-01 \\
                        CoRoT-6 &   7 &   5 &   7 &      1.0028e-01 \\
                     Kepler-13A &   5 &   3 &   5 &      1.3753e-01 \\
                      Kepler-17 &   9 &   8 &   1 &      3.2176e-04 \\
                      Kepler-63 &   5 &   3 &   5 &      1.0420e-02 \\
                       HAT-P-11 &   7 &   6 &   1 &      8.3935e-02 \\
                         $\tau$ Boo &   3 &   1 &   1 &      4.0068e-01 \\
                       WASP-107 &   3 &   3 &   1 &      2.0495e-01 \\
                          AU~Mic &   7 &   4 &   7 &      3.5363e-02 \\
\hline
\end{tabular}
\label{table_tidal_numbers}
\end{center}
\end{table}
%%%%%%%%%%%%%%%%%%%%%%%%%%%%%%%%%%%%%

%%%%%%%%%%%%%%%%%
%%%%%%%%%%%%%%%%%%%%%%%%%%%%%%%%%%%%%
%%%%%%%%%%%%%%%%%
%%%%%%%%%%%%%%%%%%%%%%%%%%%%%%%%%%%%%
%%%%%%%%%%%%%%%%%
%%%%%%%%%%%%%%%%%%%%%%%%%%%%%%%%%%%%%
%%%%%%%%%%%%%%%%%
%%%%%%%%%%%%%%%%%%%%%%%%%%%%%%%%%%%%%
%%%%%%%%%%%%%%%%%
%%%%%%%%%%%%%%%%%%%%%%%%%%%%%%%%%%%%%
%%%%%%%%%%%%%%%%%
%%%%%%%%%%%%%%%%%%%%%%%%%%%%%%%%%%%%%
%%%%%%%%%%%%%%%%%
%%%%%%%%%%%%%%%%%%%%%%%%%%%%%%%%%%%%%
%%%%%%%%%%%%%%%%%
%%%%%%%%%%%%%%%%%%%%%%%%%%%%%%%%%%%%%
%%%%%%%%%%%%%%%%%
%%%%%%%%%%%%%%%%%%%%%%%%%%%%%%%%%%%%%
%%%%%%%%%%%%%%%%%

The parameters of  the investigated systems as computed from our model are listed in Tables~\ref{table_param_sys} and~\ref{table_param_sys1}. In Table~\ref{table_param_sys}, from the left to the right, we list the name of the system; the mass $M$ and the radius $R$ of the adopted MESA stellar model for the target; the ratio $r_{0}/R$; the parameter $\sigma$; the pressure $p_{0}$ and the density $\rho_{0}$ at $r_{0}$ in the stellar model; and the minimum and maximum of the characteristic timescale $\tau$ that depend on the range of the magnetic field $B$ (cf. Eq.~\ref{tau_defin}). In Table~\ref{table_param_sys1}, from the left to the right, we list the name of the system;  the minimum and maximum values of the parameter $\beta$, of the ratio of the nondimensional Alfven frequency $\tilde{\omega}_{\rm A} = \sqrt{2}\, m f$ to the inertial frequency $\tilde{\omega}_{\rm I} = 2\, \Omega_{\rm s} \tau$, and of the Alfven velocity $v_{\rm A}$, all three depending on the magnetic field $B$; the minimum timescale of growth of the oscillation amplitude $\Delta t$ as given by Eq.~(\ref{ampl_time}) for $\xi = 10^{6}$~m and $\delta =-2\times 10^{-6}$;  the width of the resonance $\Delta \omega_{\rm res}/\Omega_{\rm s}$ as given by Eq.~(\ref{resonance_width});  the minimum tidal timescale $\tau_{\Omega_{\rm s}}$ as given by Eqs.~(\ref{tidal_torque}) and (\ref{dOmegadt}) at resonance; and the maximum value of $\xi$ allowed by the turbulent damping as estimated from Eq.~(\ref{ampl_turb}). The tidal timescale is estimated for a stellar moment of inertia $I= 6\times 10^{46}$~kg~m$^{2}$ and assuming that the tidal power is given by Eq.~(\ref{res_power}) with $\zeta = 10^{-14}$~s and considering a flux tube diameter $d_{\rm ft} = 5 \times 10^{6}$~m.  
%%%%%%%%%%%%%%%%%%%%%%%%%%%%%%%%%%%%%%%%%%%
\begin{table*}
\caption{Parameters of our model for the analyzed systems (first part).}
\begin{small}
\begin{tabular}{cccccccccccccccccc}
\hline
System & $M$ & $R$ & $r_{0}/R$ & $f$ & $\sigma$ & $\Omega_{\rm s}$ & $p_{0}$ & $\rho_{0}$ & $g_{0}$ & $\tau_{\rm min}$ & $\tau_{\rm max}$ \\
                                                           & (M$_{\odot}$) & (R$_{\odot}$) & & & & (s$^{-1}$) & (Pa) & (kg~m$^{-3}$) & m~s$^{-2}$ & (s) & (s)  \\ 
                                                                                                                      \hline
         CoRoT-2 &   1.00 &   0.89 &  0.726 &  0.112 &   -1.811 &  1.598e-05 &   1.06e+13 &   3.29e+02 &   6.40e+02 &  1.451e+07 &  1.451e+09 \\
         CoRoT-4 &   1.15 &   1.21 &  0.805 &  0.077 &   -1.962 &  7.903e-06 &   3.91e+11 &   2.26e+01 &   3.32e+02 &  3.931e+06 &  3.931e+08 \\
%         CoRoT-6 &   1.05 &   0.95 &  0.744 &  0.103 &   -1.866 &  1.145e-05 &   5.33e+12 &   1.86e+02 &   5.64e+02 &  1.098e+07 &  1.098e+09 \\ % weak field range 
         CoRoT-6$^{*}$ &   1.05 &   0.95 &  0.744 &  0.103 &   -1.866 &  1.145e-05 &   5.33e+12 &   1.86e+02 &   5.64e+02 &  1.569e+05 &  1.098e+09 \\ % strong field range 
%      Kepler-13A &   1.70 &   1.79 &  0.120 &  0.717 &    0.525 &  6.952e-05 &   5.84e+15 &   3.31e+04 &   1.64e+03 &  3.101e+08 &  3.101e+10 \\
       Kepler-13A$^{*}$ &   1.70 &   1.79 &  0.120 &  0.717 &    0.525 &  6.952e-05 &   5.84e+15 &   3.31e+04 &   1.64e+03 &  4.430e+06 &  3.101e+10 \\
       Kepler-17 &   1.15 &   1.08 &  0.795 &  0.081 &   -1.945 &  6.116e-06 &   9.58e+11 &   4.66e+01 &   4.22e+02 &  5.271e+06 &  5.271e+08 \\
%       Kepler-63 &   1.00 &   0.89 &  0.726 &  0.112 &   -1.811 &  1.346e-05 &   1.06e+13 &   3.29e+02 &   6.40e+02 &  1.451e+07 &  1.451e+09 \\ % weak field range
       Kepler-63$^{*}$ &   1.00 &   0.89 &  0.726 &  0.112 &   -1.811 &  1.346e-05 &   1.06e+13 &   3.29e+02 &   6.40e+02 &  2.073e+05 &  1.451e+09 \\  % strong field range 
        HAT-P-11 &   0.80 &   0.79 &  0.675 &  0.136 &   -1.674 &  2.384e-06 &   2.85e+13 &   7.78e+02 &   7.23e+02 &  2.238e+07 &  2.238e+09 \\
          $\tau$~Boo &   1.40 &   1.40 &  0.868 &  0.039 &   -1.998 &  2.197e-05 &   7.39e+09 &   8.65e-01 &   2.61e+02 &  4.831e+05 &  4.831e+07 \\
        WASP-107 &   0.70 &   0.65 &  0.673 &  0.141 &   -1.502 &  4.253e-06 &   6.91e+13 &   1.79e+03 &   8.98e+02 &  2.885e+07 &  2.885e+09 \\
           AU~Mic &   0.50 &   0.79 &  0.303 &  0.953 &    0.634 &  1.504e-05 &   3.39e+14 &   6.36e+03 &   3.35e+02 &  2.007e+08 &  2.007e+10 \\
         \hline
\end{tabular}
Note. Systems marked with an asterisk have an explored range of the magnetic field strength between $10$~G and $7 \times 10^{4}$~G. 
\end{small}
%\caption{Parameters of our model for the analyzed system (first part).}
\label{table_param_sys}
\end{table*}
%%%%%%%%%%%%%%%%%%%%%%%%%%%%%%%%%%%%%%%%%%
%%%%%%%%%%%%%%%%%%%%%%%%%%%%%%%%%%%%%%%%%%%
\begin{table*}
\caption{Parameters of our model for the analyzed systems (second part).}
\begin{small}
\begin{tabular}{cccccccccccccccccc}
\hline
               System & $\beta_{\rm min}$ & $\beta_{\rm max}$ &  $(\tilde{\omega}_{\rm A}/\tilde{\omega}_{\rm I})_{\rm min}$ & $(\tilde{\omega}_{\rm A}/\tilde{\omega}_{\rm I})_{ \rm max}$ & $v_{\rm alf \, min}$ & $v_{\rm alf \, max}$  & $\Delta t_{\rm min}$ &  $\Delta \omega_{\rm res}/\Omega_{\rm s}$ &  $\tau_{\Omega_{\rm s}}$ & $\xi_{\rm max}$ \\ 
                                                                                                                         &      & & & & (m~s$^{-1}$) & (m~s$^{-1}$) & (yr) &  & (yr) & (m) \\
 \hline
% wrong parameters                         CoRoT-2 &        2.67e+09 &        2.67e+13 &        1.71e-05 &        1.71e-03 &        4.92e-02 &        4.92e+00 &        1.83e+03 &       4.647e-11 &        3.38e+05 &        5.58e+11 \\
                CoRoT-2 &     2.67e+09 &        2.67e+13 &        2.74e-05 &        2.74e-03 &        4.92e-02 &        4.92e+00 &        5.58e+07 &       4.647e-11 &        8.08e+14 &        4.46e+06 \\                         
                        CoRoT-4 &        9.82e+07 &        9.82e+11 &        1.75e-05 &        1.75e-03 &        1.88e-01 &        1.88e+01 &        1.19e+04 &       1.117e-10 &        1.28e+08 &        1.16e+11 \\
%                        CoRoT-6 &        1.34e+09 &        1.34e+13 &        2.90e-05 &        2.90e-03 &        6.54e-02 &        6.54e+00 &        1.57e+09 &       7.447e-11 &        1.06e+18 &        1.46e+05 \\
                        CoRoT-6$^{*}$ &        2.73e+05 &        1.34e+13 &        2.90e-05 &        1.88e-01 &        6.54e-02 &        4.58e+02 &        1.03e+07 &       3.914e-11 &        1.06e+18 &        1.46e+05 \\
 %                    Kepler-13A &        1.47e+12 &        1.47e+16 &        7.06e-07 &        7.06e-05 &        4.91e-03 &        4.91e-01 &        3.05e+08 &       8.318e-12 &        4.59e+11 &        2.18e+10 \\
                      Kepler-13A$^{*}$ &        2.99e+08 &        1.47e+16 &        7.06e-07 &        4.93e-03 &        4.91e-03 &        3.43e+01 &        4.37e+06 &       8.276e-12 &        4.59e+11 &        2.18e+10 \\
                      Kepler-17 &        2.41e+08 &        2.41e+12 &        1.42e-04 &        1.42e-02 &        1.31e-01 &        1.31e+01 &        4.99e+06 &       1.999e-10 &        5.29e+14 &        5.02e+05 \\
%                      Kepler-63 &        2.67e+09 &        2.67e+13 &        1.22e-05 &        1.22e-03 &        4.92e-02 &        4.92e+00 &        4.51e+08 &       6.008e-11 &        9.97e+15 &        6.37e+06 \\
                      Kepler-63$^{*}$ &        5.45e+05 &        2.67e+13 &        1.22e-05 &        8.08e-02 &        4.92e-02 &        3.44e+02 &        2.42e+06 &       3.263e-11 &        9.97e+15 &        6.37e+06 \\
                       HAT-P-11 &        7.15e+09 &        7.15e+13 &        1.08e-04 &        1.08e-02 &        3.20e-02 &        3.20e+00 &        9.90e+10 &       1.140e-09 &        5.63e+20 &        2.45e+02 \\
                         $\tau$~Boo &        1.86e+06 &        1.86e+10 &        2.58e-05 &        2.57e-03 &        9.59e-01 &        9.59e+01 &        6.82e+01 &       2.538e-11 &        1.95e+05 &        5.09e+12 \\
                       WASP-107 &        1.74e+10 &        1.74e+14 &        2.43e-05 &        2.43e-03 &        2.11e-02 &        2.11e+00 &        3.54e+04 &       5.355e-10 &        6.59e+06 &        2.27e+10 \\
                          AU~Mic &        8.51e+10 &        8.51e+14 &        8.94e-06 &        8.93e-04 &        1.12e-02 &        1.12e+00 &        1.28e+14 &       4.003e-11 &        6.24e+25 &        1.03e+02 \\
\hline
\end{tabular}
Note. Systems marked with an asterisk have an explored range of the magnetic field strength between $10$~G and $7 \times 10^{4}$~G. 
\end{small}
\label{table_param_sys1}
\end{table*}
%%%%%%%%%%%%%%%%%%%%%%%%%%%%%%%%%%%%%%%%%%

For Kepler-13A, we assume that the flux tube is located at the outer boundary of the inner convection zone at $r_{0} = 0.2154$~R$_{\odot}$ because this early-type star has an outer radiative zone.   {The rotational evolution of this star can be different from that of the lower main-sequence hosts. Its present fast rotation could imply a stronger surface dynamo action and a magneto-centrifugal acceleration of its wind \citep[e.g.,][]{Johnstoneetal21}. Those effects can lead to a faster evolution of its spin that could drive it away from the present spin-orbit resonance over a timescale probably shorter than those of the other considered systems (cf. Sect.~\ref{discussion}).} 

AU~Mic is also peculiar because it is in the pre-main-sequence phase of its evolution and has a deep outer convection zone with $r_{0}/R = 0.303$. As a consequence, the values of the other parameters are remarkably different from those of the other systems having main-sequence hosts. In particular, the slender flux tube approximation is not well verified in the case of AU~Mic because $H$ is comparable with the assumed radius $r_{0}$ of the unperturbed flux tube.

Considering the results in Tables~\ref{table_param_sys} and~\ref{table_param_sys1}, the hypotheses required for the validity of the magnetostrophic approximation are well verified in all the cases because $\beta \gg 1$, the Alfven velocity $v_{\rm A}$ is much smaller than the rotation velocity $\Omega_{\rm s} r_{0}$, $\tilde{\omega}_{\rm A}/\tilde{\omega}_{\rm I} \ll 1$, and $\omega_{\rm res}/\Omega_{\rm s} \ll 1$ \citep[cf. Sect.~5.2 of][]{FerrizMasSchussler94}. The maximum oscillation amplitude $\xi_{\rm max}$ allowed by turbulent dissipation is comparable with or larger than the assumed amplitude of $10^{6}$~m, except in the cases of HAT-P-11 and AU~Mic whose spin-orbit commensurability cannot be explained with the present version of our model (see below). 

Looking at the results in Figs.~\ref{corot2_plot} to \ref{fig_aumic_plot_a}, we see that the proposed model yields a close commensurability for all the systems when a magnetic field strength between 10 and 1000~G is assumed,  except for Kepler-17 and HAT-P-11 for which a commensurability within 1\% requires a field smaller than $\sim 650$~G or $\sim 500$~G, respectively. {Slender flux tubes with such fields can be held stable inside the overshoot layer by the action of the convective downdrafts (see Sect.~\ref{internal_m_field}).} For CoRoT-6, Kepler-13A, and Kepler-63, we find that a field stronger than $10^{3}$~G is required to have an excitation timescale of the forced oscillations $\Delta t_{\rm min} \la 10^{7}$~yr as for the other systems. This requires $B \sim 7 \times 10^{4}$~G for CoRoT-6 and $B \sim 4 \times 10^{4}$~G for Kepler-63 and Kepler-13A, respectively, which implies a less tight spin-orbit commensurability for CoRoT-6 and Kepler-63 as the resonant frequency can increase up to $\sim 0.08 \, \Omega_{\rm s}$ and $\sim 0.03 \, \Omega_{\rm s}$ for these two systems, respectively. The tension with the observations could be alleviated by assuming a subadiabatic gradient $|\delta | < 2 \times 10^{-6}$ (cf. Figs.~\ref{corot6_plot} and~\ref{kepler63_plot}). 
In all the cases, smaller values of the field $B$ produce a closer commensurability because $\omega_{\rm res}$ becomes smaller.  We verified that even the flux tubes with the strongest assumed fields (up to $ B \sim 7 \times 10^{4}$~G in CoRoT-6) are stable against the undulatory instability according to Eq.~(\ref{ns_stability_condition}) for all values of $m$, so that they can be stably stored inside the overshoot layers of their respective stars, {provided that they have diameters smaller than $\sim 10^{5}$~m to be held down by the action of the convective downdrafts (cf. Eq.~\ref{eq_downdraft}). }

The only systems with an excitation timescale $\Delta t_{\rm min}$ much longer than $10^{7}$ yr are HAT-P-11 and AU~Mic that have also an extremely long tidal timescale. Both the very long $\Delta t_{\rm min}$ and $\tau_{\Omega_{\rm s}}$ are due to the very small tidal potential that is a consequence of the large $a/R$ ratio and small planetary mass in those systems. Therefore, our mechanism cannot account for their spin-orbit commensurability, at least in its present version. It could be that even extremely small tidal effects could be amplified by the operation of the stellar hydromagnetic dynamo as conjectured by, for example, \citet{Stefanietal19}, but we postpone the consideration of alternative mechanisms or any extension of the present model to future works. We only note that, once excited by an unspecified mechanism, the resonant oscillations have good chances to be observed in HAT-P-11 and AU~Mic thanks to the reduced wind-braking mechanism envisaged in Sect.~\ref{maintaining_osc} because it slows down the evolution of stellar rotation, while the tidal evolution of the planetary orbit is negligible owing to their small planetary mass and large $a/R$ ratio. 

Concerning the consequential mechanisms that can co-operate to keep the systems close to resonance, we see that the tidal timescale $\tau_{\Omega_{\rm s}}$ is shorter than $\approx 10^{8}$ yr only for CoRoT-4, $\tau$ Boo, and WASP-107. All the other targets have much longer tidal timescales despite the resonant oscillations, which in several cases are much longer than the main-sequence lifetimes of their host stars. In such systems, only the mechanism invoking a reduced or stalled rotational evolution can work. In those systems, the tidal evolution of the planetary orbit is slow because $a/R$ is rather large and tidal inertial waves cannot be excited in the convective zones of their stars with the exception of the synchronous systems CoRoT-4 and $\tau$~Boo and the cases where $P_{\rm rot} \leq 2 P_{\rm orb}$ \citep[e.g.,][]{Ahuiretal21}. Therefore, only the reduced wind braking can cooperate to maintain the resonance condition for long time intervals in most of our systems.

\section{Discussion and conclusions}
\label{discussion}
We have introduced a mechanism to explain the close commensurability between stellar rotation and orbital mean motion observed in several systems with close-by planets. It is based on the tidal resonant excitation of oscillations in a toroidal magnetic flux tube with a field $ \la 10^{3}-10^{4}$~G  stored inside the stellar overshoot layer beneath their convection zone. The tidal potential responsible for the excitation of the oscillations must have a very low frequency in the reference frame rotating with the star, which accounts for the close commensurability and allows us to treat the problem in the magnetostrophic approximation. The extremely weak intensity of the tidal potential is compensated for by the very long time interval ($\approx 10^{7}$ yr) available to amplify the oscillations thanks to the very low viscosity and magnetic diffusivity of the stellar plasma in the overshoot layer. We propose that a self-regulating mechanism keeps a system very close to the resonance peak, as required to have a significant amplification of the oscillations, because of the extreme sharpness of the resonance itself. 

We find that our mechanism may work in eight of the ten considered systems, while for the other two the tidal potential is too small to excite the oscillations during the main-sequence lifetime of the host star. Nevertheless, when such oscillations are excited, they may  strongly influence the evolution of the stellar rotation by producing an additional resonant tidal torque or by decreasing the rate of angular momentum loss caused by the stellar wind. Both mechanisms can remarkably increase the time needed to cross the resonance, thus increasing the probability of observing a system in a resonant state, that is, in a close spin-orbit commensurability. 

{The long-term stability of the resonant flux tubes is a crucial requirement of our model. From the values of the minimum amplification time, $\Delta t_{\rm min}$, listed in Table~\ref{table_param_sys1}, we see that a stability timescale as short as $\sim 10^{4}$ years can still be viable for the systems with the strongest tidal interactions, which make up $\sim 30$\% of our sample, but in any case the weakness of the tidal potential requires a stability timescale much longer than typical stellar activity cycles for the amplification of the oscillations.  As a consequence, an extension of our model parameter space to flux tube fields $\ga 10^{4}-10^{5}$~G would make the flux tubes subject to the undulatory instability, the timescale of which is too short ($\sim 1$~yr) to allow for any significant amplification of the oscillations by the tidal potential. Moreover, a stronger field will increase the resonant frequency, $\omega_{\rm res}$, making the spin-orbit commensurability less tight because the frequency of the exciting tidal potential, $\omega_{mn}$, must become larger to match $\omega_{\rm res}$.  }

It is instructive to consider the ratio of the magnetic buoyancy $B^{2}/(2\mu H)$ to the tidal force $\rho_{0} \nabla \Psi_{lmn}$ per unit volume in our systems, where the buoyancy force is considered because of the crucial role it plays in the flux tube stability and oscillation dynamics and because it is also representative, in order of magnitude, of the other forces acting on the flux tube, notably the magnetic tension. %\LEt{ Verify that your intended meaning has not been changed.} 
This ratio is listed in Table~\ref{table_buoy_over_tide} for a magnetic field of the resonant flux tube ranging from $10$ to $10^{3}$~G, that is, the typical range predicted by our model. The ratio varies widely among the different systems because of the different planet masses, different orbit semimajor axes, and different $l,m,n$ numbers determining the harmonic component of the potential that can excite resonant waves in the flux tube. We see that our model can work best when that force ratio is smaller than or comparable with unity, while it is less successful in systems such as HAT-P-11 or AU~Mic, where the force ratio is much larger than unity for all the considered  field strengths. In these cases, the large ratio is due to the weakness of the tidal force that is exciting the resonant oscillations. 

%%%%%%%%%%%%%%%%%%%%%%%%%
\begin{table}
\caption{Ratio of buoyancy, $F_{\rm B}$, to tidal force,  $F_{\rm T}$, in our systems.}
\begin{center}
\begin{tabular}{ccc}       
\hline                                                                                   
                      System  & $(F_{\rm B}/F_{\rm T})_{\rm min}$  & $(F_{\rm B}/F_{\rm T})_{\rm max}$\\
                      \hline 
                       CoRoT-2  &         2.95e+01  &         2.95e+05 \\
                        CoRoT-4  &         1.08e-01  &         1.08e+03   \\
                        CoRoT-6  &         2.07e+03  &         2.07e+07   \\
                     Kepler-13A  &         4.37e-03  &         4.37e+01   \\
                      Kepler-17  &         1.98e+02  &         1.98e+06   \\
                      Kepler-63  &         1.23e+02  &         1.23e+06   \\
                       HAT-P-11  &         9.74e+04  &         9.74e+08   \\
                         $\tau$~Boo  &         1.37e-02  &         1.37e+02   \\
                       WASP-107  &         4.16e-03  &         4.16e+01   \\
                          AU~Mic  &         3.84e+05  &         3.84e+09   \\                    
\hline
\end{tabular}
\label{table_buoy_over_tide}
\end{center}
\end{table}
%%%%%%%%%%%%%%%%%%%%%%%%%%%%%%%%%%%%%

The possibility of keeping a system in resonance for time intervals comparable with its main-sequence lifetime is restricted to stars with close-by massive hot Jupiters thanks to the remarkable tidal torque  {{or reduced angular momentum loss rate in the winds induced}} by the resonant oscillations in those cases. {An interesting example is provided by CoRoT-2, which has a distant visual main-sequence companion (CoRoT-2B) whose age has been constrained to be $\ga 5$~Gyr from the lack of any detectable X-ray emission \citep{PoppenhaegerWolk14}.  Therefore, there is a considerable tension with the fast rotation of the primary CoRoT-2A, which is indicative of an age $\la 0.3-0.6$~Gyr according to the rotational evolution observed in single stars of the same spectral type \citep[e.g.,][]{Johnstoneetal21}. In such a case, we can speculate that the close-by massive planet CoRoT-2b is responsible for the prolonged fast rotation of its host star, having kept it locked into a {{8:3}} spin-orbit resonance for gigayears. 

From a more general point of view, considering the force ratio listed in Table~\ref{table_buoy_over_tide}, one could propose it as a qualitative indicator of the possibility of exciting resonant oscillations leading to a persistent spin-orbit commensurability, but such an application is hampered by our ignorance of the dependence of the strength of the flux tube field on the stellar evolutionary phase. Our basic assumption that the resonant flux tubes are very long-lived makes it difficult to specify their connection to the stellar dynamo field, which is usually assumed to oscillate on timescales of a few decades. For this reason, while we can specify how the dynamo is affected by the evolution of stellar rotation, this is presently not possible for the fields of the resonant magnetic flux tubes. Their connection to the stellar dynamo and to stellar magnetism in general is a point left for future investigations. }

The tidal interaction produced by the resonant flux tubes considered in our model can be classified as a particular case of dynamical tide that requires a very low tidal frequency in the stellar reference frame. Therefore, its range of frequency operation is very limited in comparison to, for example, the dynamical  tide produced by inertial waves that are excited when the tidal frequency falls in the range $[-2\Omega_{\rm s}, 2\Omega_{\rm s}]$ \citep{OgilvieLin07}. Nevertheless, our mechanism producing an enhanced tidal interaction can be relevant for systems very close to synchronization and can explain how CoRoT-4 and $\tau$~Boo can maintain a synchronous rotation state, which is difficult to interpret with other tidal models \citep[cf.][]{Borsaetal15}. 

An additional effect of the tidal torque produced by a planet in spin-orbit resonance on an oblique orbit is the damping of the obliquity itself. The damping is significant for a high obliquity but becomes smaller and smaller as the obliquity decreases to values $\la 20^{\circ}-30^{\circ}$ because of the rapid decrease of the coefficient $A_{lmn}$ (cf. Fig.~\ref{almn_plot}). We postpone the investigation of the evolution of the obliquity to future work because it is outside the scope of the present paper. Additionally, it requires a dedicated analysis in view of a possible application to the interpretation of the observations that show a statistical decrease in obliquity with a decrease in the effective temperature of the host star \citep[see][for a recent review]{Andersonetal21}. 

The reduced or stalled rotational evolution that occurs as a consequence of the oscillations invoked by our model  can significantly affect the application of gyrochronology to stars that host close-by planets. This point has already been presented in the literature \citep[e.g.,][]{Pont09,Lanza10,Brown14,Maxtedetal15}, and the present investigation offers a physical mechanism in support of it. 

It is interesting to consider the chromospheric counterparts of the photospheric starspots rotating in close commensurability with the planet that form according to our model (cf. Sect.~\ref{secondary_reg_mechanisms}). These chromospheric spots will show the same rotational behavior and the same azimuthal wavenumber of the resonant oscillations.  Therefore, they cannot account for the observations of hot spots synchronized with the orbital motion of a close-by planet proposed by \citet{Shkolniketal05,Shkolniketal08}, because they are characterized by an $m=1$ pattern and the ratio $P_{\rm rot}/P_{\rm orb}$ is not a rational number. An exception is the case of synchronous systems with $m=1$, such as $\tau$~Boo.  Indeed, the deviating behavior of the hot spot of $\tau$~Boo suggested by \citet{Cauleyetal19} may point to a different origin than a star-planet magnetic interaction and can be interpreted in the framework of the present model as being due to the emergence of flux tubes  perturbed by the $m=1$ resonant oscillation mode. 

Finally, we note that the possibility of a close commensurability with the stellar rotation period being twice the orbital period may be relevant for the internal heating of ultra-short-period planets around very active young stars, as recently proposed by \citet{Lanza21}. 

\begin{acknowledgements}
The author is grateful to an anonymous Referee for a careful reading of the manuscript and several comments that allowed him a remarkable improvement of this work. 
This investigation has been supported by the PRIN INAF 2019 "Planetary systems at young ages (PLATEA)", the PI of which is Dr.~S.~Desidera. The use of the MESA-Web interface in October 2021 to compute interior stellar models is also acknowledged. AFL thanks Dr. Daniela Domina for her kind librarian assistance in retrieving a copy of some of the cited articles. 
\end{acknowledgements}

\appendix

\section{Tidal potential produced by a planet on a circular oblique orbit}
\label{potential_development}

We considered a point-like planet of mass $m_{\rm p}$ on a circular orbit of radius $a$, inclined by an angle $i$ with respect to the equator of the host star; the orbital mean motion is $\Omega_{0}$. We derived the tidal potential inside the host star following the approach of \citet{BravinerOgilvie15}. 

We adopted a nonrotating reference frame with the origin at the barycenter of the star and the polar axis along the orbital angular momentum, $\vec L$. The spin angular momentum of the star is $\vec S$, and the obliquity, $i$, is the angle between $\vec L$ and $\vec S $. In the adopted reference frame, the polar coordinates of a point are  $(r, \theta^{\prime\prime}, \phi^{\prime\prime})$ and the planet position vector is ${\vec r}_{\rm p} = (a, \pi/2, \Omega_{0} t)$, where $t$ is the time. 

The gravitational potential $\Phi ({\vec r})$ produced by the planet at a point $\vec r$ inside the star is
\begin{equation}
\Phi ({\vec r}) = \frac{Gm_{\rm p}}{| {\vec r}_{\rm p} - {\vec r}|} =  \frac{Gm_{\rm p}}{a} \sum_{l=0}^{\infty} \left( \frac{r}{a} \right)^{l} P_{l} (\cos \gamma),
\end{equation}
where we have developed the potential in a series of Legendre polynomials $P_{l}$ of degree $l$ and $\cos \gamma = \hat{{\vec r}}_{\rm p} \cdot \hat{{\vec r}}$ with the hat here indicating a unit vector. 

Applying the addition theorem of the spherical harmonics, we express the Legendre polynomials in terms of the spherical harmonics in our reference frame yielding
\begin{equation}
\Phi   =   \frac{Gm_{\rm p}}{a} 
\sum_{l=0}^{\infty} \frac{4\pi}{2l+1} \left(\frac{r}{a} \right)^{l} \sum_{n=-l}^{l} (-1)^{n} Y_{l}^{-n} \left(\frac{\pi}{2}, 0 \right) Y_{l}^{n}(\theta^{\prime\prime}, \phi^{\prime\prime}) e^{-jn\Omega_{0} t},  
\label{a2}
\end{equation}
where $n-l$ must be even in order to have  $Y_{l}^{-n}\left( \frac{\pi}{2}, 0 \right) \propto P_{l}^{-n} (0) \not= 0$. 

Next we considered another nonrotating spherical polar coordinate system $(r, \theta,  \phi)$ with the polar axis $\theta=0$ along the stellar spin $\vec S$. It is obtained from the $(r, \theta^{\prime\prime}, \phi^{\prime\prime})$ system by making a rotation by an angle $i$ around the axis $\hat{\vec S} \times \hat{\vec L}$. The  spherical harmonics  in these two coordinate systems are related by the formula \citep[e.g.,][]{MorrisonParker87}
\begin{equation}
Y_{l}^{n} \left( \theta^{\prime\prime}, \phi^{\prime\prime} \right) = \sum_{m=-l}^{l} Y_{l}^{m} \left( \theta, \phi \right) d_{mn}^{l} (i), 
\label{a3}
\end{equation}
where the elements of the Wigner $\vec d$ matrix are functions of the obliquity $i$ and are given by
\begin{equation}
d_{mn}^{l} (i) = \left[ \frac{(l+n)! (l-n)!}{(l+m)! (l-m)!} \right]^{1/2} \left( \sin \frac{i}{2} \right)^{n-m} \left( \cos \frac{i}{2} \right)^{n+m} P^{(n-m, n+m)}_{l-n} (\cos i), 
\label{a4}
\end{equation}
where $P^{(\alpha, \beta)}_{k} (x)$ is a Jacobi polynomial with  $\alpha$, $\beta$, and $k>0$  integers. They are defined as
\begin{equation}
P^{(\alpha, \beta)}_{k} (x) \equiv \sum_{s=0}^{k} \left( 
\begin{array}{c} k+\alpha \\ 
k-s 
\end{array} \right) 
\left( 
\begin{array}{c} k+\beta \\ 
s 
\end{array} \right) \left(\frac{x-1}{2} \right)^{s} \left( \frac{x+1}{2} \right)^{k-s}
,\end{equation}
with $-1\leq x \leq 1,$ and 
\begin{equation}
\left( \begin{array}{c} z  \\ 
w  \end{array} \right) = \frac{z!}{w! (z-w)!},
\end{equation} 
with $z \geq w$, $z, w \in \mathbb{N}$ and $0! =1$ by definition. 

Substituting Eq.~(\ref{a3}) into Eq.~(\ref{a2}), exchanging the order of the summations over $n$ and $m$, and considering a reference frame rotating with the stellar angular velocity $\Omega_{\rm s}$, so that $\phi \rightarrow \phi + \Omega_{\rm s} t$, we finally find 
\begin{eqnarray}
\Phi =   \frac{Gm_{\rm p}}{a} 
\sum_{l=0}^{\infty} \sum_{m=-l}^{l} \sum_{n=-l}^{l} (-1)^{n} \frac{4\pi}{2l+1} Y_{l}^{-n} \left( \frac{\pi}{2}, 0 \right) d_{mn}^{l}(i) \left( \frac{r}{a}\right)^{l} \nonumber \\ 
\times \, Y_{l}^{m} \left( \theta, \phi \right) e^{j\hat{\omega}_{mn} t}, 
\label{a5}
\end{eqnarray}
where the tidal frequency  $\hat{\omega}_{mn} = m\Omega_{\rm s} - n\Omega_{0}$ and $l-n$ must be even. The expression of the coefficients $A_{lmn}$ given in Eq.~(\ref{amnl_eq}) immediately follows from Eq.~(\ref{a5}), where we remind that the components of the tidal potential are those with $l \geq 2$. 
%To adhere to the definition of the tidal frequency introduced in Sect.~\ref{tides}, we need to switch the sign of $m$ and $n$ in the present definition of $\omega_{mn}$ which does not produce any physical change to the system because both integers range between $-l$ and $l$. [no longer needed because I changed the definition of \omega_{mn} in Sect. 3.1.]

\end{document}